\documentclass[prx,twocolumn,superscriptaddress,nofootinbib,longbibliography]{revtex4-1}
\usepackage{amsfonts}
\usepackage{graphicx}
\usepackage{bm}
\usepackage{amssymb}
\usepackage{color}
\usepackage{amsmath}
\usepackage{amstext}
\usepackage{latexsym}
\usepackage[usenames,dvipsnames]{xcolor}
\usepackage[colorlinks=true,citecolor=Cerulean,linkcolor=RubineRed,urlcolor=Cerulean]{hyperref}

\newcommand\ba{\begin{eqnarray}}
\newcommand\ea{\end{eqnarray}}
\newcommand\be{\begin{equation}}
\newcommand\ee{\end{equation}}

\def\non{\nonumber}

\begin{document}
\title{Universal finite-time thermodynamics of many-body quantum machines\\ from Kibble-Zurek scaling}
\author{Revathy B. S}
\affiliation{Department of Physics,  Indian Institute of Technology Palakkad,  Palakkad, 678557,  India}
\author{Victor Mukherjee}
\affiliation{Department of Physical Sciences, IISER Berhampur, Berhampur 760010, India}
\author{Uma Divakaran}
\affiliation{Department of Physics,  Indian Institute of Technology Palakkad,  Palakkad, 678557,  India}

\author{Adolfo del Campo}
\affiliation{Donostia International Physics Center,  E-20018 San Sebasti\'an, Spain}
\affiliation{IKERBASQUE, Basque Foundation for Science, E-48013 Bilbao, Spain}
\affiliation{Department of Physics, University of Massachusetts, Boston, MA
02125, USA}
\affiliation{Theory Division, Los Alamos National Laboratory, MS-B213, Los Alamos, NM 87545, USA}
\date{\today}

\begin{abstract}

We demonstrate the existence of  universal features in the finite-time thermodynamics 
of quantum machines by considering a many-body quantum Otto cycle  in which the working medium is
driven across quantum critical points during the unitary strokes. 
Specifically, we consider  a quantum engine  powered by dissipative energizing and 
relaxing baths. We show that under very generic conditions, the output work 
is governed by the  Kibble-Zurek 
mechanism, i.e., it exhibits a universal power-law scaling with  the driving speed  through the critical points.
We also optimize the finite-time thermodynamics as a function of the driving speed.  The 
 maximum power and the corresponding efficiency take a universal form, and are reached for an optimal speed that is  governed by the critical exponents.
We exemplify our results by considering a transverse-field Ising spin chain as the working medium.
For this model, we also show how the  efficiency and power vary as the 
engine becomes critical.

\end{abstract}


\maketitle

\section{Introduction}

Advances in quantum science and technology have made possible the laboratory implementation of minimal quantum devices such as heat engines and refrigerators using a variety of platforms that include   trapped ions \cite{Rossnagel16,Maslennikov17,vonLindenfels19}, nitrogen vacancy centers  \cite{Klatzow19}, and nuclear magnetic resonance experiments \cite{Peterson18}.
Quantum engines (QE)  transform heat and possibly other resources into some kind of useful work \cite{gemmer2009quantum}. 
Their study paves the way to identify quantum effects in their performance.  In particular, one may wonder whether there exist scenarios exhibiting a quantum advantage with no 
classical counterpart \cite{Jaramillo16,Klatzow19,mukherjee19anti}.

To a large extent, the study of quantum engines has been 
restricted to single-particle systems \cite{Kosloff17}. 
Such devices already display 
nontrivial features when their operation involves quantum 
synchronization \cite{noufal20quantum}, non-thermal coherent 
and squeezed  reservoirs 
\cite{scully03extracting,rossnage14nanoscale,
Gardas15,Niedenzu18},
quantum measurements 
\cite{Elouard17,Elouard17b,Cottet17} and quantum metrology  \cite{brunner17, bhattacharjee20quantum},
in the presence of  quantum coherence over sustained many 
cycles \cite{watanabe17quantum}, 
or in the small action limit, when different cycles 
become thermodynamically equivalent \cite{Uzdin15}.

Quantum thermal machines  with many-body working mediums (WMs) 
may allow us to harness many-body effects, such as entanglement and  other quantum 
correlations 
for operation with enhanced power and efficiency \cite{Jaramillo16}.
Shortcuts to adiabaticity have been shown to enhance the performance
of many body quantum  thermal machines \cite{hartmann19}.
Quantum statistics can boost the performance of Szilard engines \cite{Kim11,Reimann18}.  Similarly, the performance of quantum Otto Cycles in both the adiabatic \cite{Zheng15} and 
finite-time operation \cite{Jaramillo16} can exhibit  an enhancement due to bosonic quantum statistics, while a detrimental one has been predicted in the fermionic case. 
Other many-particle effects that can be harnessed for the engineering of QE include super-radiance \cite{Hardal15} and many-body localization \cite{halpern2019quantum}, 
while novel configurations become feasible, e.g., by using spin networks \cite{Turkpence17}.
Many-particle QE are also required for scalability and the possibility of suppressing quantum 
friction during their finite-time operation \cite{Deng13,delcampo14,Beau16,Funo17,chenu18thermodynamics}
which has been explored in the laboratory with trapped Fermi gases \cite{Shujin18,Diao18}.

Quantum criticality may offer new avenues to boost the performance of heat engines, 
as a result of the diverging length and time-scales close to a phase transition  \cite{sachdev99quantum}. 
The enhancement of microscopic fluctuations to approach Carnot efficiency in finite time was 
proposed in \cite{Polettini15}.
Further, the scaling theory of second-order phase transitions  has been used to show that 
the ratio  between the output power and the deviation of the efficiency  from the Carnot limit can be optimized at criticality \cite{campisi16the}. 
In adiabatic interaction-driven heat engines, quantum criticality has also shown to 
optimize the output power \cite{Yang19}.

In this work, we introduce a quantum  Otto cycle with a working medium 
that exhibits a quantum phase transition.
In particular, we consider the family of free-fermionic models that include 
paradigmatic instances of critical spin systems such as the quantum Ising and XY chains, 
as well as higher dimensional models. 
As a result, our setting is of direct relevance to current efforts for building many-particle QEs, 
with e.g., trapped ions. We explore  how signatures of universality in the critical dynamics of the working medium 
carry over the finite-time thermodynamics of the heat engine.

Remarkably, we show that the scaling of the work output of such QEs with the driving time follows a universal power law resulting from the
Kibble-Zurek mechanism. This result paves the way for
the hitherto 
 unexplored field of universal finite-time thermodynamics describing quantum machines driven through quantum critical points.
To the best of our knowledge, such a connection has not been
explored before, and is the focus of our paper.

In Sec. \ref{secII}, we introduce the model of a many body 
Otto cycle using a free-Fermionic WM. We discuss Kibble-Zurek scaling and its connection to the output work and power of quantum Otto cycles in Section \ref{seckzm}, while Section \ref{secIIB}
introduces an efficiency bound depending on dynamical critical exponent.
We focus on the particular example of a transverse Ising spin chain WM in Sec. \ref{secIII}
which is further divided into two subsections depending upon the different phases 
the WM explores during unitary strokes. We also provide analytical 
expressions
for the energies exchanged in each stroke and compare them
with numerics.
Finally we conclude in Sec. \ref{secIV}.

\section{Many body Otto Cycle}
\label{secII}

The use of spins as WM opens a wide range of opportunities recognized early 
on \cite{Geva92,Quan07}. Recent experiments have implemented 
single-spin quantum heat engine \cite{Peterson18,vonLindenfels19} and test fluctuation theorems 
in single strokes \cite{Serra14,Smith18}.  WM composed of interacting spins 
such as multiferroics have been proposed  \cite{Azimi14,Chotorlishvili16} whereas it is shown that WMs
with cooperative effects boost engine properties \cite{NiedenzuKurizki18}.
Quantum critical spin systems in quantum thermodynamics have also been considered under adiabatic performance \cite{Cakmak16,Ma17},  shortcuts to adiabaticity \cite{Cakmak19} and the limit of sudden driving \cite{Dorner12,Nigro19}. Such settings preclude the study of signatures of universality  associated with the quantum critical dynamics in the finite-time protocols, which is our focus. 

We consider an Otto cycle with a many-body WM, described by the Hamiltonian
\begin{eqnarray}
H&=& \sum_k \Psi_k^{\dagger} \tilde{H}_k \Psi_k\non,\\
\tilde{H}_k&=&(\lambda+a_k) \sigma^z + b_k \sigma^+ + b_k^* \sigma^-,
\label{eq_genHmat}
\end{eqnarray}
with $\sigma^+ = (\sigma^x + i \sigma^y)/2$, $\sigma^-=(\sigma^x-i \sigma^y)/2$, and
$\sigma^x$, $\sigma^y$, $\sigma^z$ being the usual Pauli matrices.
Here $\tilde{H}_k$ is a $2\times 2$ matrix in  a basis given by 
$\Psi_k^{\dagger}= (c_{1k}^{\dagger}, c_{2k}^{\dagger})$ where $c_{jk},~c_{jk}^{\dagger}$ ($j = 1,2$)
are fermionic operators for the $k$-th momentum mode. 
Such a Hamiltonian 
includes  widely studied models, such as the transverse-field Ising and XY chains \cite{bunder99effect, lieb61two, pfeuty70the, dziarmaga10dynamics, dutta15quantum},
and the two dimensional Kitaev model \cite{kitaev06anyons, chen08exact, sengupta08exact}, 
through suitable choices of $\lambda,~a_k$ and $b_k$. 
This Hamiltonian exhibits a quantum critical point (QCP) at $\lambda = \lambda_{\rm c}$, 
when the energy gap 
$\Delta=2 \sqrt{(\lambda_{\rm c}+a_{k})^2 + |b_{k}|^2}$ between the ground state and first excited state
vanishes, for the critical mode $k = k_{\rm c}$.
The density matrix of such a system can be written
in a basis consisting of $|0,0\rangle$, $|1_{1k},0\rangle$, $|0,1_{2k}\rangle$, $|1_{1k},1_{2k}\rangle$
where the first index corresponds to presence (1) or absence (0) of $c_{1k}$ fermion.
Similarly, the second index corresponds to $c_{2k}$ fermions.
 It is to be noted that the unitary dynamics generated by the Hamiltonian $\tilde H_k$
mixes $|1_{1k},0\rangle$, and $|0,1_{2k}\rangle$ only.
 As we shall see later, the non-unitary dynamics
allows mixing along the other two basis too \cite{keck17dissipation, bandyopadhyay18exploring}. 
We denote the full $4 \times 4$ Hamiltonian matrix by $H_k$.

Before dwelling on the dynamics in Fourier space,
let us briefly discuss  its real space counterpart.
One of the prominent instances within the family of Hamiltonians in 
 Eq. (\ref{eq_genHmat})  is that of the
Ising  and the XY models in a transverse field (we assume  a ring geometry) which takes the real-space form
\ba
H &=& -\sum_i M_i \left(c_i^{\dagger}c_{i+1} - c_ic^{\dagger}_{i+1} \right) \non\\
&+& N_i\left(c_i^{\dagger}c_{i+1}^{\dagger} - c_ic_{i+1}\right) + R_i \left(c_i^{\dagger}c_{i} - c_ic_{i}^{\dagger}\right).
\label{hamilfer}
\ea
Here $i$ denotes the site index, $c_i, c_i^{\dagger}$ are Fermionic 
annihilation and creation operators, respectively, and $M_i, ~N_i,~R_i$ are scalars \cite{lieb61two}. 
Such a Hamiltonian can be generated, for example, using a WM consisting of 
interacting-Fermions in an optical lattice setup \cite{schreiber15observation}.
If $M_i$, $N_i$ and $R_i$ are site independent, one can perform Fourier transform 
of the Hamiltonian to express it in the  form of Eq. (\ref{eq_genHmat}).
We shall discuss more on this Hamiltonian in Sec. \ref{secIII}.

The quantum Otto cycle alternates between unitary and nonunitary strokes.  
We now describe below the four general stages of the Otto cycle in 
details (see Fig. \ref{fig_cycle}):

\begin{figure}[h]
\includegraphics[width = 0.71\columnwidth]{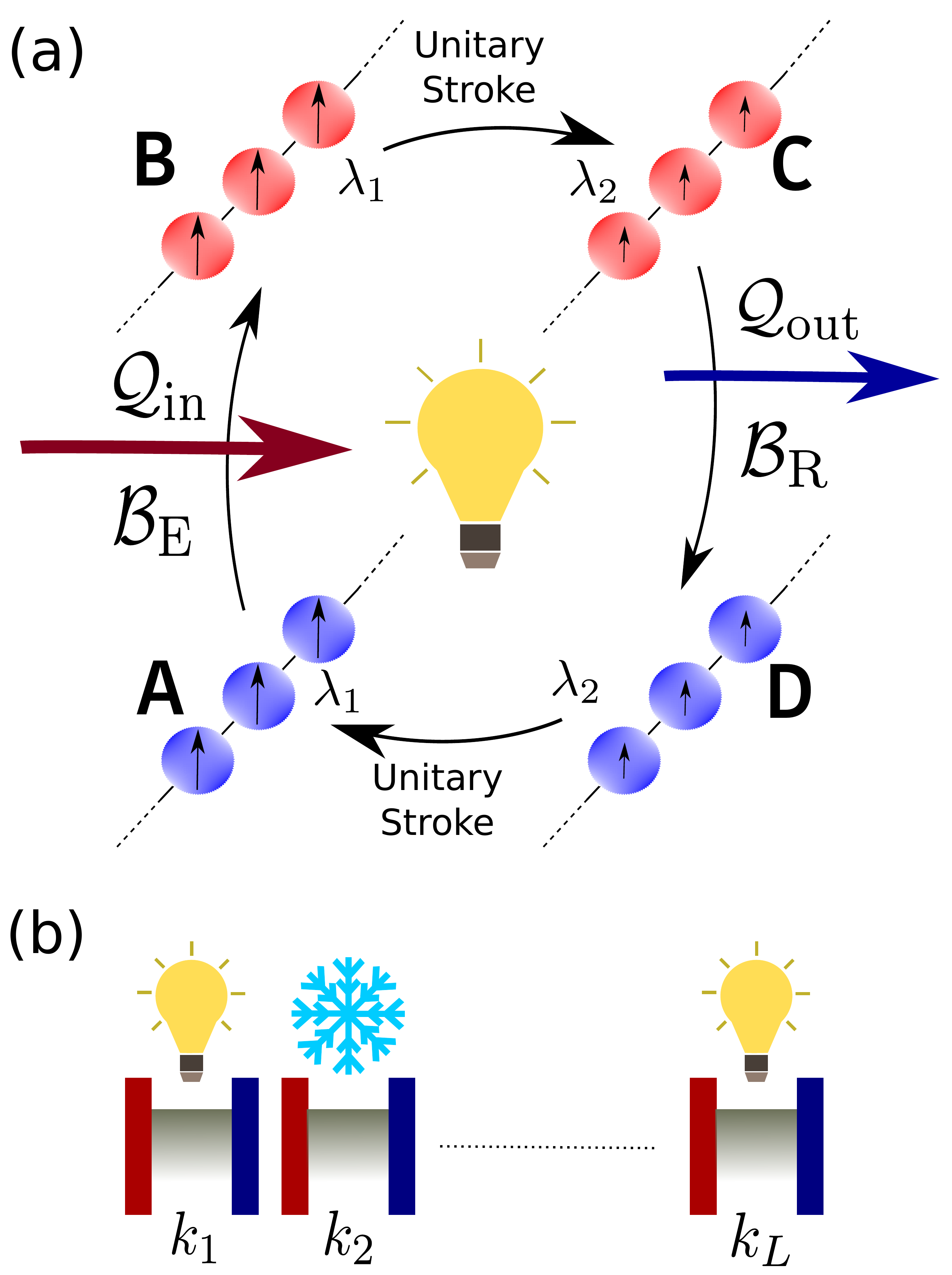}
\caption{{\bf Schematic diagram of a many-body quantum Otto cycle} (a) Schematic diagram of a quantum Otto cycle with a many-body system as the working medium. We get a net output work in the heat engine regime (shown by the glowing light bulb). (b) In the equivalent 
momentum space, the interacting many-body QE can be represented by independent quantum thermal machines corresponding to the different decoupled Fermionic $k$ modes, 
each acting as a heat engine (shown by glowing light bulbs), or as a refrigerator (shown by the snow-flake),
or even as a heat distributor (not shown here).}
\label{fig_cycle}
\end{figure}

\begin{enumerate}
\item {\it{Stroke 1}} (${\bf A} \to {\bf B}$): 
The WM is subjected to a constant Hamiltonian (Eq. \eqref{eq_genHmat}) 
with $\lambda=\lambda_1$, while being coupled 
to a dissipative \emph{energizing} bath $\mathcal{B}_{\rm E}$ for a 
time $\tau_{\rm E}$ as shown in Fig. \ref{fig_cycle}a, thus resulting in non-unitary dynamics. 
We assume $\tau_{\rm E}$ to be large enough so that the WM reaches
 the steady state.

In general the dissipative dynamics undergone by the density matrix $\rho(t)$
is given by
\begin{eqnarray}
\frac{d\rho}{dt}=-i[H, \rho] + \mathcal{D}[\rho]
\label{eq_lindblad}
\end{eqnarray}
with $\hbar$ set to unity, and
$\mathcal{D}[\rho]$ is the non-unitary part of the dynamics
generated due to the interaction of the system with the bath. The exact form of 
$\mathcal{D}[\rho]$ depends upon the nature of the bath and its interaction with the system. 
Here we consider  baths with unique steady states. This can be achieved, for example, by coupling the WM to a thermal bath at a finite temperature.

Alternatively, one can consider Markovian baths
coupled locally to the Fermionic modes shown in 
Eq. \eqref{hamilfer}, with $\mathcal{D}[\rho]$ given by
\ba
\mathcal{D}\left[\rho \right] &=& \sum_i \tilde{\kappa}_i \left(L_i\rho L_i^{\dagger} - \frac{1}{2}\{\rho, L_i^{\dagger}L_i \} \right).
\ea 
Here $\tilde{\kappa}_i$ is related to system-bath coupling strength 
for the site $i$,  $L_i$ are local Lindblad operators that describe the interaction of the
Fermion at site $i$ with the bath. For $L_i=c_i$ (see Eq. (\ref{hamilfer})) 
and site independent $\tilde{\kappa}_i$,
it can be shown that the Fourier transform  of $\mathcal{D}[\rho]$ does not mix
different modes, so that we arrive at mode-dependent non-interacting local baths 
in the free-Fermionic representation in momentum space. 
The existence of non-interacting Fermionic modes implies the state 
$\rho(t)$ of the many-body WM can be written as 
$\rho(t)=\bigotimes_k \rho_k(t)$, with the time-evolution of $\rho_k(t)$
given by the differential equation \cite{keck17dissipation, bandyopadhyay18exploring}
\begin{eqnarray} 
\frac{d\rho_{k}}{dt} &=& -i[H_{k}, \rho_{k}]  + \mathcal{D}_k[\rho_k]\non\\
\mathcal{D}_k[\rho_k] &=&  \kappa_{1}^E \left(c_{1k} \rho_{k}
c_{1k}^{\dagger} -\frac{1}{2}\lbrace c_{1k}^{\dagger}c_{1k}, 
\rho_k \rbrace \right) \non\\
&+& \kappa_2^{\rm E}\left(c_{1k}^{\dagger} \rho_{k} c_{1k} -\frac{1}{2}\lbrace c_{1k}
c_{1k}^{\dagger},\rho_k \rbrace \right)  \non\\ 
&+& \kappa_{3}^E \left(c_{2k} \rho_{k}
c_{2k}^{\dagger} -\frac{1}{2}\lbrace c_{2k}^{\dagger}c_{2k}, \rho_k \rbrace \right) \non\\
&+& \kappa_4^{\rm E} \left(c_{2k}^{\dagger} \rho_{k} c_{2k} -\frac{1}{2}\lbrace c_{2k}
c_{2k}^{\dagger},\rho_k \rbrace \right) 
\label{eq_stroke1}
\end{eqnarray}

 Here $\kappa_j^{\rm E}$ ($j = 1,2,3,4$)
are positive constants related to the \emph{energizing} bath, which depend on the coupling strength 
between the WM and the bath.
The energy exchanged in this stroke is denoted as $\mathcal{Q}_{\rm in}$.

\item {\it{Stroke 2}} (${\bf B} \to {\bf C}$): The system is decoupled from the bath 
at ${\bf B}$ and $\lambda$ is varied linearly in time as $t/\tau_1$ from 
$ \lambda_1$ (at ${\bf B}$)  to $\lambda_2$ 
(at ${\bf C}$)  in a time interval $\tau_1$, such that the WM undergoes a unitary 
dynamics described by
\begin{eqnarray}
\frac{d\rho_k}{dt}=-i [H_k,\rho_k].
\label{eq_unitary}
\end{eqnarray}
We consider $\lambda_1>\lambda_2$ in this paper.
Work is done on or by the system in this stroke.

\item {\it Stroke 3} (${\bf C} \to {\bf D}$): The WM is now coupled to 
a \emph{relaxing} bath $\mathcal{B}_{\rm R}$ at {\bf C} of Fig. \ref{fig_cycle}a, 
for a time duration $\tau_{\rm R}$, at a constant $\lambda = \lambda_2$. 
The evolution equation will be similar to that given in Eq. (\ref{eq_stroke1})
with appropriate couplings $\kappa_1^{\rm R}\dots\kappa_4^{\rm R}$ related to $\mathcal{B}_{\rm R}$.
Quantum critical dynamics are more pronounced for systems close to their ground states. 
Consequently, universal scaling behavior is to be expected by considering 
a \emph{relaxing} bath which takes the WM to its ground state in this stroke. 
In principle, we can tune the relaxing bath coupling parameters 
 such that it either takes the system
to its ground state or to some steady state corresponding to the bath parameters. 
We denote the energy exchanged in this stroke with $\mathcal{Q}_{\rm out}$.

\item {\it Stroke 4} (${\bf D} \to {\bf A}$): The system is decoupled 
from $\mathcal{B}_{\rm R}$  and $\lambda_2$ (at ${\bf  D}$)
is varied back to $\lambda_1$ (at ${\bf A}$) linearly in a time interval  $\tau_2$ as $t/\tau_2$. 
Once again, work is done on or by the system in this stroke.

One can operate the QE in a steady state cycle, by repeating the above described cycle.
It is to be noted that depending upon the values of $\lambda_1$ and $\lambda_2$, we may or may not 
cross the critical point. We consider both of these possibilities in this paper.
\end{enumerate}

At the end of any stroke, the energy of the system is calculated using
\begin{eqnarray}
	\mathcal{E} = \text{Tr}(H \rho)=\sum_k \text{Tr} (H_k \rho_k).
\label{eq_energy}
\end{eqnarray}
We choose the parameters 
$\kappa_1^{\rm E}, \dots, \kappa_4^{\rm E}, \kappa_1^{\rm R}, \dots, \kappa_4^{\rm R}$, $\lambda_1$ and $\lambda_2$
such that energy $\mathcal{Q}_{\rm in}$ is absorbed when coupled to $\mathcal{B}_{\rm E}$ in {\it {stroke}} 1, 
while a smaller amount $\mathcal{Q}_{\rm out}$ is released when coupled to $\mathcal{B}_{\rm R}$ in {\it {stroke}} 3, 
such that the setup operates as a heat engine, with a net output work 
$\mathcal{W} = -\left(\mathcal{Q}_{\rm in} + \mathcal{Q}_{\rm out}\right)$. 
We assume the following sign convention for the energy flows: $\mathcal{Q}_{\rm in}$, $\mathcal{Q}_{\rm out}, \mathcal{W}$
are positive (negative) if the WM energy increases (decreases). 
For the Otto cycle to operate as a heat engine, we need $\mathcal{Q}_{\rm in} > 0$, $\mathcal{Q}_{\rm out} < 0, \mathcal{W} < 0$.  
On the other hand, $\mathcal{Q}_{\rm in} < 0,~\mathcal{Q}_{\rm out} > 0,~\mathcal{W}>0$ corresponds to a refrigerator, 
and $\mathcal{Q}_{\rm out} <0,~\mathcal{W} > 0$ denotes a heat distributor \cite{mukherjee16speed}. 
We characterize the performance of the heat engine in terms of its efficiency $\eta$ 
\begin{eqnarray}
\eta=\frac{\mathcal{Q}_{\rm in}+\mathcal{Q}_{\rm out}}{\mathcal{Q}_{\rm in}} = -\frac{\mathcal{W}}{\mathcal{Q}_{\rm in}}
\label{eq_efficiency}
\end{eqnarray}
as well as the power output $\mathcal{P}$
\ba
\mathcal{P}=-\frac{\mathcal{Q}_{\rm in}+\mathcal{Q}_{\rm out}}{\tau_{\rm total}},
\label{eq_power}
\end{eqnarray}
where $\tau_{\rm total} =  \tau_{\rm E} + \tau_{\rm R} +\tau_1 + \tau_2$ being the total cycle time.
For a WM that can be described in terms of non-interacting momentum modes as 
shown in Eq. \eqref{eq_genHmat}, it follows that
\ba
\mathcal{Q}_{\rm in} = \sum_k \mathcal{Q}_{\rm in}(k);~~~ \mathcal{Q}_{\rm out} = \sum_k \mathcal{Q}_{\rm out}(k),
\ea
where $\mathcal{Q}_{\rm in}(k),~\mathcal{Q}_{\rm out}(k)$ denote the energy flows corresponding to the $k$-th mode.
We note that even if the complete setup acts as a QE, the individual fermionic modes 
may act as QE, refrigerator, or heat distributor, depending on the details of the operation 
and WM, see Fig. \ref{fig_cycle}b.

\section{Universal  thermodynamics}

\subsection{Universal Kibble-Zurek scaling in output work}
\label{seckzm}

Two of the strokes of the Otto cycle perform unitary dynamics during which a quantum critical point
may be crossed depending upon $\lambda_1$ and $\lambda_2$. 
The universal dynamics in terms of excitations produced due to 
diverging relaxation time at the critical point 
is a well studied subject 
\cite{polkovnikov05universal,zurek05dynamics,mukherjee07quenching,
deffner17}, and can be explained through the adiabatic-impulse approximation \cite{damski06adiabatic}.
 Consider a system which is initially 
prepared in the ground state of a time dependent
Hamiltonian such that it crosses the critical point linearly
as $t/\tau$.
The amount of density of defects (excitations) 
with respect to the ground state corresponding to the 
Hamiltonian at final time, follows a universal power-law with the
rate of variation $1/\tau$. 
The exponent of the power-law is dependent on
the equilibrium critical exponents of the quantum 
critical point crossed. This power-law relation 
is known as Kibble-Zurek  scaling, 
after its proponents T. W. B. Kibble and W. H. Zurek,
and is given by \cite{Polkovnikov11,delcampo14universality}
\ba
n_{\rm ex} \sim \tau^{-\frac{\nu d}{\nu z+1}},
\label{kzex}
\ea
where $n_{\rm ex}$ denotes the density of excitations, $d$ is the dimensionality of 
the system and $\nu$, $z$  are the correlation length and dynamical critical exponents,  respectively.
The density of excitations $n_{\rm ex}$ in turn gives rise to the excitation energy 
$E_{\rm ex}$, i.e., the energy of the system above the instantaneous ground state, 
which can also be expected to scale with the rate of quench $1/ \tau$ 
\cite{Polkovnikov11,caneva07adiabatic, keck17dissipation}.  Signatures of Kibble-Zurek mechanism have also been verified experimentally in transverse-field Ising model, using trapped ions \cite{cui2016experimental} and  a quantum annealer \cite{Bando20}.

These universal signatures may govern quantum Otto engines under
 the following very generic conditions:
\begin{itemize}
\item The \emph{relaxing} bath $\mathcal{B}_{\rm R}$ takes the WM close to its ground state.
This is one of the important conditions in order to arrive at the scaling derived below.
\item The WM is driven at a finite rate across a quantum critical point (or points) 
during the unitary stroke {\bf D} $\to$ {\bf A},  i.e., $\tau_2$ is finite.
\item The \emph{energizing} bath $\mathcal{B}_{\rm E}$ takes the WM to a 
unique steady state with high entropy.
\end{itemize}

The first condition of the relaxing bath $\mathcal{B}_{\rm R}$ taking the WM close to its ground state can be realized for example by considering $\mathcal{B}_{\rm R}$ to be a cold thermal bath  at temperature $T_{\rm c}$ much smaller 
than the energy scale $\mathcal{E}_{\rm WM}$ associated with the WM, where $\mathcal{E}_{\rm WM} \sim L^{-z}$ for $L \gg |\lambda - \lambda_c|^{-\nu}$
and $\mathcal{E}_{\rm WM} \sim |\lambda - \lambda_{\rm c}|^{\nu z}$ for  $L \ll |\lambda - \lambda_c|^{-\nu}$ \cite{sachdev99quantum}. Similarly, one can realize the last condition of the WM being in a high-entropy state at {\bf B} 
by considering a thermal \emph{energizing} 
bath $\mathcal{B}_{\rm E}$ with temperature $T_{\rm h} \gg \mathcal{E}_{\rm WM}$, such that the corresponding 
steady state of the WM is 
close to a maximum entropy state, which in general is the unique state with all the energy levels equally populated. Therefore for the practical scenario of a WM with finite $L$, and therefore finite $\mathcal{E}_{\rm WM}$, finite values of $T_{\rm c}$ and $T_{\rm h}$ would suffice, as long as the above conditions are met.
The unitary stroke {\bf B} $\to$ {\bf C} cannot change the entropy of the WM, i.e., all the energy levels need to be equally populated at {\bf C} as well, in order to preserve the entropy.
Consequently  the states of the WM at {\bf C} and {\bf B} remain 
approximately equal, for any value of $\tau_1$. We note that,  
the states of the WM at {\bf C} and {\bf B} can also be 
approximated 
to be equal  if the WM is quenched rapidly across the quantum critical 
point during the unitary stroke {\bf B} $\to$ {\bf C}, i.e., $\tau_1 \to 0$, 
for any form of $\mathcal{B}_{\rm E}$ or of the state of the WM at {\bf B}.
This is needed in order to write an expression
for work done which is only related to the excitations
in the stroke D to A as discussed below.

The work done is given by
\ba
\mathcal{W} &=& - \left(\mathcal{Q}_{\rm in} + \mathcal{Q}_{\rm out}\right) \non\\
\mathcal{Q}_{\rm in}  &=& \mathcal{E}_{\rm B} - \mathcal{E}_{\rm A} = \mathcal{E}_{\rm B} - 
\mathcal{E}_{\rm A}^{\rm G} - \mathcal{E}_{\rm ex, A} \non\\
\mathcal{Q}_{\rm out} &=& \mathcal{E}_{\rm D}^{\rm G} - \mathcal{E}_{\rm C},
\label{WQs}
\ea
where $\mathcal{E}_{\rm A}$, $\mathcal{E}_{\rm B}$ and $\mathcal{E}_{\rm C}$ are the energies of the 
WM at {\bf A}, {\bf B} and {\bf C}, respectively, $\mathcal{E}_{\rm A}^{\rm G}$ and 
$\mathcal{E}_{\rm D}^{\rm G}$ are the ground state energies of the WM at {\bf A} and {\bf D}, respectively, 
while $\mathcal{E}_{\rm ex, A}$ denotes the excitation energy of the WM at {\bf A}. 
The implementation of the  engine ensures that $\mathcal{E}_{\rm A}^{\rm G}$, $\mathcal{E}_{\rm B}$, 
$\mathcal{E}_{\rm C}$ and $\mathcal{E}_{\rm D}^{\rm G}$ are independent of $\tau_2$, 
while the Kibble-Zurek mechanism  manifests itself through the presence of $\mathcal{E}_{\rm ex, A}$ 
in the output work:
\ba
\mathcal{W} - \mathcal{W}_{\infty} = \mathcal{E}_{\rm ex, A}.
\label{workkzm}
\ea
Here $\mathcal{W}_{\infty} = -\left(\mathcal{E}_{\rm B} - \mathcal{E}_{\rm A}^{\rm G} + \mathcal{E}_{\rm D}^{\rm G}  - \mathcal{E}_{\rm C}\right)$ 
is the work output in the limit $\tau_2 \to \infty$, which depends only on $\lambda_1$, $\lambda_2$, 
and the steady-state of the bath $\mathcal{B}_{\rm E}$. Remarkably, as seen above (Eq. \eqref{workkzm}), 
the output work shows the same scaling with $\tau_2$ as the excess energy, upto an additive constant. 
For a quench that ends at the critical point, one arrives at a 
universal scaling relation
\cite{grandi10quench,fei2020work}
\ba
\mathcal{E}_{\rm ex, A} \sim \tau_2^{-\frac{\nu(d+z)}{\nu z+1}}\\
{\rm or,~}\mathcal{W} - \mathcal{W}_{\infty} \sim  \tau_2^{-\frac{\nu (d+z)}{\nu z+1}}.
\label{workkzm3}
\ea
By contrast, for quenches across the critical point the excess energy is not universal in general.  Yet, for systems and quench protocols in which the excess energy is
proportional to the density of defects, such as the examples we consider below,
the scaling  \eqref{workkzm3} is modified as
\ba
\mathcal{W} - \mathcal{W}_{\infty} \sim  \tau_2^{-\frac{\nu d}{\nu z+1}}.
\label{workkzm2}
\ea

The above results, \eqref{workkzm}, \eqref{workkzm3},
and \eqref{workkzm2},
are the highlights of our paper. They establish
a connection between Kibble-Zurek mechanism, which has been
traditionally studied in the context of cosmology
\cite{kibble80some, zurek85cosmological, zurek96osmological}
and quantum phase transitions in closed quantum systems
\cite{zurek05dynamics,polkovnikov05universal,dutta15quantum},
and the quantum thermodynamics of QE.

Universal scaling relations in systems driven 
through quantum critical points 
have been widely studied in closed quantum systems 
\cite{polkovnikov05universal, zurek05dynamics, mukherjee07quenching,  dziarmaga10dynamics,  dutta15quantum}.
However, whether such scaling forms hold 
in the presence of dissipation is  a  delicate question with 
no unique answer \cite{hoyos07effects,Patane08,Dutta16,Bando20,
wang20dissipative}. Signatures of quantum phase transitions 
arise due to vanishing energy gaps close to criticality. 
Naturally, thermal fluctuations can be expected to destroy or 
significantly affect these signatures, at any non-zero temperature 
\cite{sachdev99quantum}. This effect can be even more pronounced 
in quantum machines, that generally involve multiple unitary and 
non-unitary strokes. In this context, one can follow the design 
presented here to engineer quantum machines powered by dissipative 
baths, and that exhibit a performance governed by the universal 
Kibble-Zurek scaling, in spite of the presence of the multiple 
unitary and non-unitary strokes. This possibility is remarkable 
as signatures of quantum critical dynamics are generally suppressed 
in quantum machines not fulfilling the above constraints encoded 
in the design; for example, if the relaxing bath does not take the 
system close to its ground state, or if the steady-state of  
energizing bath depends on the state of the WM at {\bf A}. 

One can easily extend these results to QEs involving non-linear 
quenches across quantum critical points, following the results 
reported in Ref. \cite{sen08defect}. The importance of the 
Kibble-Zurek power-law scalings in the operation of quantum machines
stems from the identification of universal signatures in the
finite-time thermodynamics of critical QE, as well as 
the optimization of their performance, to which we now turn our 
discussion.

To this end, we note that in the limit of
$\tau_{\rm{total}} \approx \tau_2 \gg \tau_1, \tau_{\rm E}, \tau_{\rm R}$,
one can use \eqref{workkzm3} and \eqref{workkzm2} to derive a scaling
relation for the output power $\mathcal{P}$
\ba
\mathcal{P} = \frac{\mathcal{W}}{\tau_2} \approx \frac{\mathcal{W}_{\infty}}{\tau_2}
+ R \tau_2^{-\frac{\nu d + x\nu z + 1}{\nu z+1}}
\label{eq_powerscaling}
\ea
where $R$ is the proportionality constant.
Here $x=1$ corresponds to crossing the critical point and $x=2$
is when $\lambda_1$ is set to its critical value.
The optimal quench rate $\tau_2^{-1} = \tau_{\rm opt}^{-1}$
delivering the maximum power can be found from the condition
$\frac{\partial \mathcal{P}}{\partial \tau_2}\big|_{\tau_{\rm opt}} = 0$ which yields
\ba
 \tau_{\rm opt} &=& \left[\frac{R \left(\nu d + x\nu z + 1 \right)}{|\mathcal{W}_{\infty}| \left(\nu z + 1 \right)} \right]^{\left(\nu z +1\right)/[\nu d+(x-1)\nu z]},
\label{tau_opt}
\ea
with the corresponding efficiency $\hat{\eta}$ at maximum power being
\ba
\hat{\eta} &=& -\frac{\mathcal{W}_{\infty} +  \mathcal{E}_{\rm ex, A}( \tau_{\rm opt})}{\mathcal{E}_{\rm B} - \mathcal{E}_{\rm A}^{\rm G} - \mathcal{E}_{\rm ex, A}(\tau_{\rm opt})}.
\label{eq_opteff}
\ea
The presence of $\mathcal{E}_{\rm ex}$ in $\mathcal{W}$
as well as in $\mathcal{Q}_{\rm in}$ renders the
corresponding efficiency $\eta$ independent of $\tau_2$, 
for large $\tau_2$ as shown clearly in section \ref{secIII}.

Furthermore, one can use \eqref{workkzm3}-\eqref{eq_opteff}
to design optimally performing
many-body quantum machines operated close to
criticality, by judiciously choosing WMs with appropriate
critical exponents and dimensionality.
For example, as one can see from \eqref{workkzm2},
other factors remaining constant,
enhancement of output work would require choosing a WM with large
dimension $d$.

The net output work
$\mathcal{W}$ might involve Kibble-Zurek
scaling arising  due to the passage from {\bf B} to {\bf C} as well, for example, if the WM remains close to its ground state
at {\bf B} and $\tau_1$ is finite.
In addition, the universal
scalings in Eqs. \eqref{workkzm3} and \eqref{workkzm2} would be
modified in the case of sudden quenches \cite{grandi10quench},
or in presence of disorder \cite{caneva07adiabatic}.

We shall later exemplify the novel results \eqref{workkzm3} 
and \eqref{workkzm2} with the
transverse Ising model as a working medium which has a well studied
quantum critical point.

\subsection{Efficiency bound}
\label{secIIB}

One can  arrive at a maximum efficiency bound $\eta_{\rm max}$ of the QE, 
by defining a maximum possible temperature $T_{\rm max}$ and a 
minimum possible temperature $T_{\rm min}$. We design the 
QE such that the maximum (minimum) possible 
energy gap $\Delta_{\rm max} =  \{\Delta(\lambda_1, k)\}_{\rm max}$ ($\Delta_{\rm min} =  \{\Delta(\lambda_2, k)\}_{\rm min}$) between two consecutive energy levels is realized at
$\lambda_1$ ($\lambda_2$), where the maximum (minimum) is taken over all the $k$ modes 
and energy gaps. For sufficiently large
$\lambda_1$ (i.e., $\left(\lambda_1 - \lambda_{\rm c}\right)^{\nu z} \gg k^z~ \forall~k$), 
$\Delta(\lambda_1, k)$ is independent of $k$, and
is a function of $\lambda_1$ alone. 
In analogy with a thermal bath, we define $T_{\rm max}$ through the following relation \cite{breuer02}:
\ba
\exp\left[-\Delta_{\rm max}/T_{\rm max} \right] &=& \kappa_1^{\rm E}/\kappa_2^{\rm E},\non\\
T_{\rm max} := \frac{\Delta_{\rm max}}{\ln \frac{\kappa_2^{\rm E}}{\kappa_1^{\rm E}}}.
\label{Tmax}
\ea
Similarly, one can define an analogous minimum possible temperature through the relation:
\ba
T_{\rm min} := \frac{\Delta_{\rm min}}{\ln \frac{\kappa_2^{\rm R}}{\kappa_1^{\rm R}}}.
\label{Tmin}
\ea
Here we have assumed $\kappa_1/\kappa_2=\kappa_4/\kappa_3$ for both the \emph{energizing} as well as the \emph{relaxing} bath. 

The net efficiency of the spin-chain QE is given by
\ba
\eta = \frac{\sum_k \left(\mathcal{Q}_{\rm in}(k) +  \mathcal{Q}_{\rm out}(k)\right)}{\sum_k \mathcal{Q}_{\rm in}(k)} = \frac{\sum_k \eta(k) \mathcal{Q}_{\rm in}(k)}{\sum_k \mathcal{Q}_{\rm in}(k)},
\label{effnet}
\ea
where $\eta(k)$ is the efficiency corresponding to the $k$-th mode. Therefore defining $\eta_{\rm max} = \{\eta(k)\}_{\rm max}$ we get
\ba
\eta \leq \eta_{\rm max}\frac{\sum_k \mathcal{Q}_{\rm in}(k)}{\sum_k \mathcal{Q}_{\rm in}(k)} = \eta_{\rm max}.
\label{etam}
\ea
For dissipative baths acting as thermal baths with mode dependent temperatures, the second law demands that each $\eta(k)$ should abide by the Carnot bound of maximum efficiency, with 
the temperatures of the hot and cold bath
depending on the mode $k$. Consequently, one can arrive at $\eta_{\rm max}$ through $T_{\rm max}$ and $T_{\rm min}$ defined above:
\ba
\eta_{\rm max} &=& 1 - \frac{T_{\rm min}}{T_{\rm max}} 
= 1 - \frac{\Delta_{\rm min}}{\Delta_{\rm max}}
\cdot\frac{\ln \left(\kappa_2^{\rm E}/\kappa_1^{\rm E}\right)}{\ln \left(\kappa_2^{\rm R}/\kappa_1^{\rm R}\right)}.
\label{etamax}
\ea

The minimum possible non-zero energy gap $\Delta_{\rm min}$  between two consecutive energy levels arise at the QCP (i.e., $\lambda_2 = \lambda_{\rm c}$), when it assumes the value 
\ba
\Delta_{\rm min} = \left(2\pi/L\right)^z,
\ea
for a WM with length $L$ \cite{sachdev99quantum}.
Consequently, for a QE operating between a $\lambda_1$ and $\lambda_2 = \lambda_{\rm c}$, we get
\ba
T_{\rm min} = \frac{\left(2\pi/L\right)^z}{\ln \frac{\kappa_2^{\rm R}}{\kappa_1^{\rm R}}}
\ea
and
\ba
\eta_{\rm max} = 1 - \frac{\left(2\pi/L\right)^z}{\Delta_{\rm max}}
\cdot\frac{\ln \left(\kappa_2^{\rm E}/\kappa_1^{\rm E}\right)}{\ln \left(\kappa_2^{\rm R}/\kappa_1^{\rm R}\right)}.
\label{etamacc}
\ea
As can be seen from Eq. \eqref{etamacc}, $\eta_{\rm max}$ increases with increasing system 
size $L$, thus showing a possible advantage offered by many-body quantum engines over few-body ones. 

Interestingly, as discussed above, $\eta_{\rm max}$ is maximum 
for an Otto cycle operating between a $\lambda_1$ and the QCP 
$\lambda_2 = \lambda_{\rm c}$. However, we note that $\eta_{\rm max}$ 
in general does not provide a tight bound. 
The equality in Eq. \eqref{etam} can be expected to hold only in 
the limit of a WM with mode independent energy gaps.
Furthermore, as shown in the example of Ising spin chain in presence of 
a transverse field WM below, contrary to the behavior of $\eta_{\rm max}$, 
the actual efficiency of the QE, even though bounded by Eq. \eqref{etamax},  
may peak slightly away from the QCP.\\

We note that the effective temperatures defined in Eqs. \eqref{Tmax} and \eqref{Tmin}, and consequently also the efficiency bound \eqref{etamacc}, depend crucially on the condition that the annihilation and
creation operators
$c_{jk}, c_{jk}^{\dagger}$ cause transitions between adjacent energy levels  for the $j = 1,2$ fermions, such that
 the dissipative baths act as thermal baths with mode-dependent temperatures for each mode $k$.

\section{A transverse Ising spin chain working medium}
\label{secIII}

We now exemplify the universality of critical QEs with 
the transverse Ising spin chain as the WM, and determine
the efficiency and  power close to, as well as away from criticality.
The Hamiltonian of transverse Ising  model (TIM) in spin space can be written as
\begin{eqnarray}
H=- \sum_{i=1}^L (J \sigma_i^x \sigma_{i+1}^x + h \sigma_i^z)
\label{Hising}
\end{eqnarray}
where $\sigma_i^{\alpha}$ denotes the Pauli matrix in the direction $\alpha$, acting at the site $i$,
and $L$ is the total number of sites or length of the system.
Without any loss of generality, we set $J$ to unity.
The Hamiltonian \eqref{Hising}, when written in terms of Jordan Wigner fermions $c_i$
followed by its Fourier transform $c_k$ can be rewritten as
\begin{eqnarray}
H &=& \sum_{k>0} \Psi_k^{\dagger} \tilde{H}_k \Psi_k,~{\rm{with}}\nonumber \\
\tilde{H}_k &=& 2(h+\cos(k)) \sigma_z + 2\sin(k) \sigma^+ + 2\sin(k) \sigma^-
\label{eq_timH}
\end{eqnarray}
~\\
where $\Psi^{\dagger}_k= (c_k^{\dagger}, c_{-k})$.
Clearly, $\lambda$ in Eq. \ref{eq_genHmat} corresponds to the transverse
field $h$, $a_k=2\cos k$ and $b_k=2\sin k$. The QCP where the gap $\Delta$
between the ground state and first excited state vanishes for this Hamiltonian
is given by $h=\pm 1$ with the critical mode $k_{\rm c}=\pi$ and $0$, respectively. 
\cite{lieb61two, pfeuty70the, bunder99effect}. 
There is a quantum phase transition from a paramagnetic phase 
for $h>1$ to a ferromagnetic phase for $h<1$ \cite{sachdev99quantum}.

Comparing Eq. (\ref{eq_genHmat}) and Eq. (\ref{eq_timH}), we find that 
$c_{1k}=c_k$, and $c_{2k}=c_{-k}^{\dagger}$.
As before, the four basis corresponds to $|0,0\rangle$, $|1_k,0\rangle$, 
$|0,1_{-k}\rangle$ and $|1_k,1_{-k}\rangle$.
The full $4\times 4$ Hamiltonian matrix $H_k$ is given by
\begin{eqnarray} H_{k} =
\begin{bmatrix} 2(h + \cos k) & 0 & 0 & 2 \sin k \\ 0 & 0 & 0 & 0\\ 0 & 0 & 0 &
0\\ 2 \sin k & 0 & 0 & -2(h + \cos k) 
\label{eq_tim4}
\end{bmatrix} 
\end{eqnarray}
such that the unitary dynamics only mixes $|0,0\rangle$ and $|k,-k\rangle$
but the non-unitary dynamics mixes the state into all four basis.
To write the evolution equation of the $4\times 4$ density matrix when connected to a bath
which is similar to Eq. (\ref{eq_stroke1}), lets choose the
interaction, and the form of the Lindblad equation as follows:
\begin{widetext}
\begin{eqnarray} 
\frac{d\rho_{k}}{dt} &=& -i[H_{k}, \rho_{k}] 
+ \left[ \mu \left(c_{k} \rho_{k}
c_{k}^{\dagger} -\frac{1}{2}\lbrace c_{k}^{\dagger}c_{k}, \rho_k\rbrace \right) 
+ \mu' \left(c_{k}^{\dagger} \rho_{k} c_{k} -\frac{1}{2}\lbrace c_{k}
c_{k}^{\dagger},\rho_k\rbrace \right) \right]\nonumber\\  
&+& \left[\mu \left(c_{-k} \rho_{k}
c_{-k}^{\dagger} -\frac{1}{2}\lbrace c_{-k}^{\dagger}c_{-k}, \rho_k\rbrace \right)
+ \mu' \left(c_{-k}^{\dagger} \rho_{k} c_{-k} -\frac{1}{2}\lbrace c_{-k}
c_{-k}^{\dagger},\rho_k\rbrace \right) \right]
\label{eq_timstroke1}
\end{eqnarray}
\end{widetext}

Eq. \eqref{eq_timstroke1} resembles that of a multilevel system coupled 
with a thermal bath, albeit with a mode dependent temperature \cite{breuer02}. 
However, the dissipative 
baths are not thermal, since they are coupled locally to the WM in the momentum space. 
In the following, we denote $\mu's$ related to \emph{energizing} bath $\mathcal{B}_{\rm E}$ with subscript E and that of \emph{relaxing} bath $\mathcal{B}_{\rm R}$ with subscript R.

The QE with TIM as the WM undergoes an Otto cycle, 
with  $\lambda_1$ ($\lambda_2$) 
replaced by $h_1$ ($h_2$).
In {\it{stroke}} 2, let $h(t)$ is changed linearly from $h_1$ to $h_2$ 
with time $t$ as $h(t) = h_1+(h_2-h_1) t/\tau_1$
 with $0<t<\tau_1$, where $\tau_1$
is related to the speed with which $h$ is varied. 
In the reverse direction during the {\it{stroke}} 4, $h$ is varied as 
$h_2+(h_1-h_2)(t-\tau_{\rm R}-\tau_1)/\tau_2$ for 
$\tau_{\rm R}+\tau_1<t< \tau_{\rm R}+\tau_1+\tau_2$. 
One can use the state $\rho_k$ and the Hamiltonian $H_k$ at the end of 
each stroke and for each $k$, to  calculate the efficiency and the power using Eq. (\ref{eq_efficiency}) and 
Eq. (\ref{eq_power}), respectively.
Depending upon the values of $h_1$ and $h_2$, the QE explores different regions of the WM phase diagram. For example, with $h_1,h_2 >1$, the WM is driven through the paramagnetic phase only, without crossing any of the critical points. When $h_1\gg1$ and $-1<h_2<1$, the 
WM crosses one critical point and explores the paramagnetic and ferromagnetic phases. On the other hand,
for $h_1\gg1$ and $h_2\ll -1$, the unitary strokes traverse the two critical points, separating 
Paramagnetic-Ferromagnetic-Paramagnetic boundaries. 

\begin{figure}
\includegraphics[width = \columnwidth]{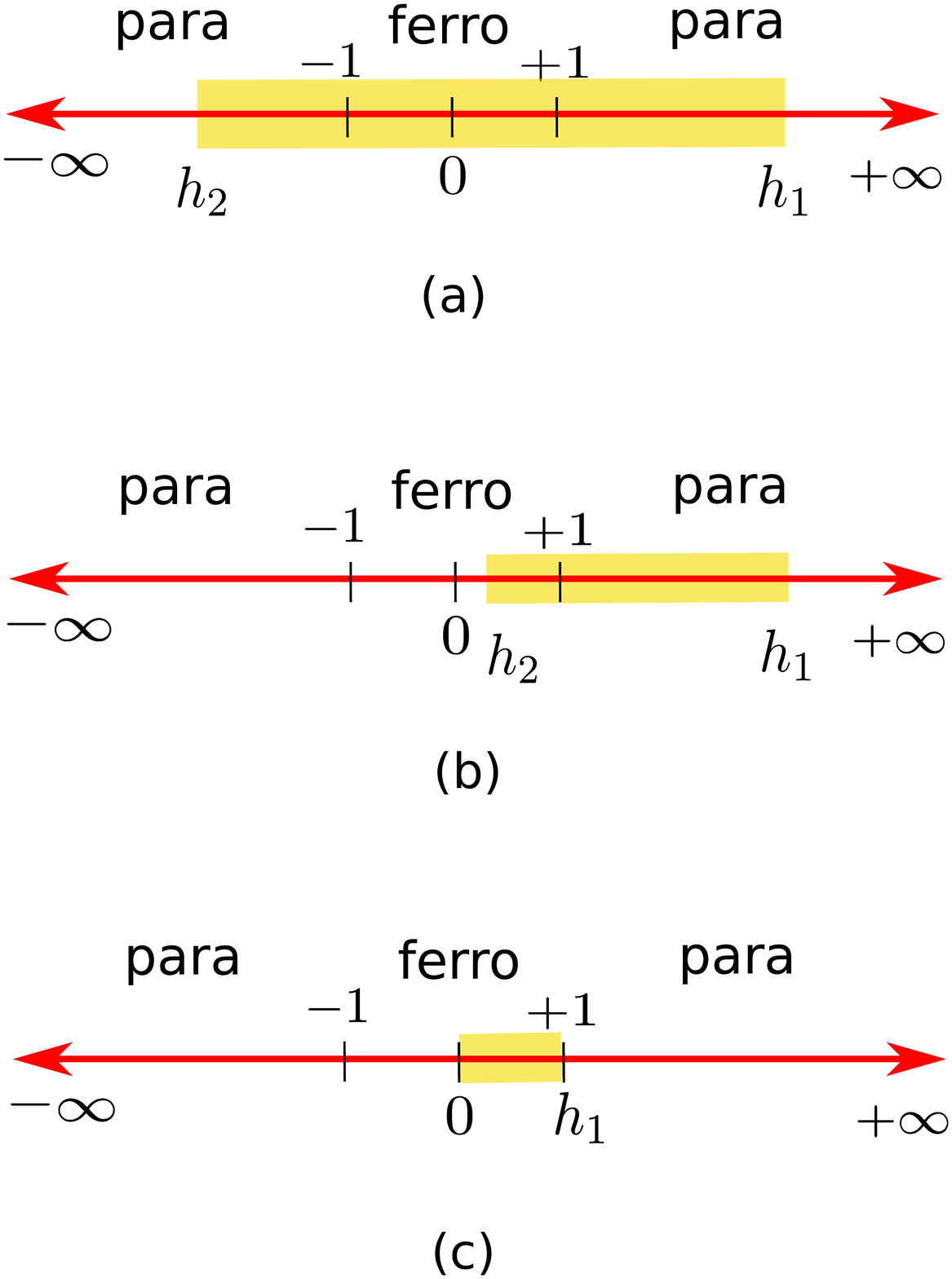}
\caption{{\bf Schematic diagram showing the different classes of QEs considered below:} (a) Para-Para QE, where the unitary strokes are between $h_1 \gg 1$ and $h_2 \ll -1$ (b) Para-Ferro QE with the unitary strokes between $h_1 \gg 1$ and $0 < h_2 < 1$ and (c) Critical-Ferro QE, where the unitary strokes are between 
$h_1 \to 1$ and $h_2 = 0$. The regimes of operations are shown by the yellow highlights. The quantum critical points at $h = \pm J = \pm 1$ separate the paramagnetic phases ($|h| > 1$) from Ferromagnetic phase ($|h| < 1$).}
\label{fig_phases}
\end{figure}

We thus consider (see Fig. \ref{fig_phases}) (i) Para-Para QE when the WM
crosses two critical points,  
(ii) Para-Ferro QE with one critical point crossed,
(iii) Critical-Ferro QE,
and (iv) Generalized QE,
and start the discussion with the engine of the first type. 
Each of these engines bring out different features as we detail next.

\subsection{Para-Para QE}

A Para-Para QE can be realized with $h_1 \gg 1$ and $h_2 \ll -1$.
The work done in a Para-Para QE admits  a 
closed form expression, 
which can directly be connected to Kibble-Zurek scaling.
In order to explore the Kibble-Zurek scaling in heat engines, 
it is important that one of the unitary dynamics
start from the ground state of the Hamiltonian. 
We choose the parameters of the \emph{relaxing} bath such that
it takes the system closest to its ground state. 
The unitary dynamics from $\bf D$ to $\bf A$ will then show the 
Kibble-Zurek scaling. 
For this, we fix $\mu_{\rm R}=1$ and $\mu_{\rm R}'=0$, since it is the $\mu$ term which
brings the system to the ground state for negative field values. 
We choose \emph{energizing} bath $\mathcal{B}_{\rm E}$ parameters as $\mu_{\rm E}<1$ and $\mu_{\rm E}'=1$.
Also, we choose $h_1 \gg 1$ and $h_2 \ll -1 $ so that 
both the critical points 
$h=\pm 1$ are crossed. The ground state for both 
the field values is paramagnetic
where $c-$ particles are also the quasiparticles.
This will help in getting  closed analytical expressions
for $\mathcal{Q}_{\rm in}$, $\mathcal{Q}_{\rm out}$ and work done, 
and finally their dependence on criticality.

\subsubsection{Analytical calculations}
\label{tim_an}

\begin{figure}
\includegraphics[height=2.5in]{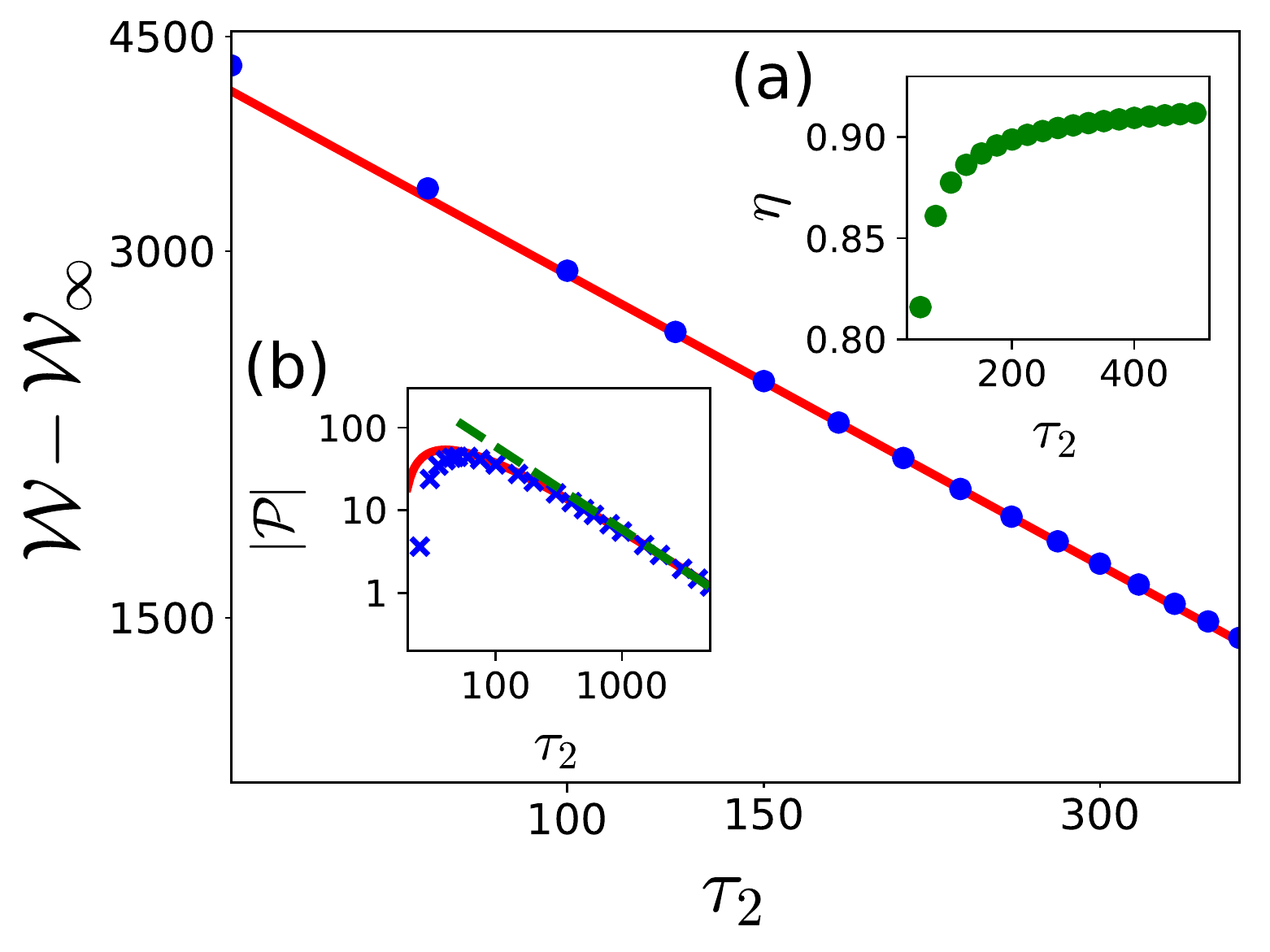}
\caption{{\bf Work output showing universal Kibble-Zurek scaling in Para-Para QE.}
The points are the numerical values and red solid line corresponds to
$\tau_2^{-1/2}$. For transverse Ising model, $d=\nu =z=1$.
Inset (a): Variation of $\eta$ with $\tau_2$.
(b) Variation of Power with
$\tau_2$. The green dashed line corresponds to $1/\tau_2$
scaling, points represent
numerical data and solid line is the analytical expression.
The parameters used are:
$L=100, h_1=70, h_2 = -5, \tau_1=0.01, \mu_{\rm E}'=1,
\mu_{\rm E}=0.995, \mu_{\rm R}'=0, \mu_{\rm R}=1$
with $\mathcal{W}_{\infty}=$-6481.205}.

\label{fig_kz}
\end{figure}

Our analytical expressions for the various energy values 
below are obtained for $h_1\gg 1$, $h_2 \ll -1$, $|h_1|\gg |h_2|$. We further consider 
 a high entropy  steady state at {\bf B} or $\tau_1$ small, or both, so that one can write the density 
matrix at $C$. We first note that the basis 
$|0,0\rangle, |1_k,0\rangle, |0,1_{-k}\rangle, 
|1_k,1_{-k}\rangle$ 
are also the eigen basis of the Hamiltonian for 
large $|h|$; see Eq. (\ref{eq_tim4}).
To calculate energies $\mathcal{E}_{\rm B}$, $\mathcal{E}_{\rm C}$, 
$\mathcal{E}_{\rm D}$ and $\mathcal{E}_{\rm A}$, 
at ${\bf B}$, ${\bf C}$, ${\bf D}$ and ${\bf A}$ 
respectively, using Eq. (\ref{eq_energy}), 
we need to write the density matrix 
at each of these points. One can see that at ${\bf B}$, when the system has reached its steady state
after connecting to the \emph{energizing} bath $\mathcal{B}_{\rm E}$ with $\mu=\mu_{\rm E}$ and $\mu'=\mu_{\rm E}'$, 
the density matrix takes the form 
\begin{equation}\label{rho_B}
\rho_{k}^{\rm B} = 
\begin{bmatrix} P_{4}^{\rm B} & 0 & 0 & 0 \\ 0 & P_{3}^{\rm B} & 0 & 0\\ 0 & 0 &
P_{2}^{\rm B} & 0\\ 0 & 0 & 0 & P_{1}^{\rm B}
\end{bmatrix},
\end{equation} 
where $P_{1}^{\rm B}, P_{2}^{\rm B},P_{3}^{\rm B}, P_{4}^{\rm B}$ are the populations in the 
energy levels $E_{1}, E_{2}, E_{3} $ and $E_{4}$ of the Hamiltonian 
with $E_{1}< E_{2}=E_{3}<E_{4}$ for $h_1 \gg 1$. Clearly, the order reverses
for $h\ll -1$. The symbol B in
superscript represents point ${\bf B}$ of the cycle. We shall use the symbol D
for quantities related to point ${\bf D}$ for similar reasons.
These probabilities can be obtained using the steady state
condition of the master equation which gives

\begin{eqnarray}\frac{P_{2}^{\rm B}}{P_{1}^{\rm B}} =
\frac{P_{3}^{\rm B}}{P_{1}^{\rm B}} = \frac{P_{4}^{\rm B}}{P_{2}^{\rm B}} =\frac{P_{4}^{\rm B}}{P_{3}^{\rm B}} = \mu_{\rm E}
\label{eq_Pratios}
\end{eqnarray} 
where $\mu_{\rm E}'=1$ as discussed before.
Also, from the normalisation condition we have
\begin{eqnarray}\label{2} P_{1}^{\rm B} + P_{2}^{\rm B} + P_{3}^{\rm B} + P_{4}^{\rm B} = 1.  
\label{eq_Pnorm}
\end{eqnarray}
From (\ref{eq_Pratios}) and (\ref{eq_Pnorm}), we get the populations in the energy levels when
connected to the ${\mathcal{B}_{\rm E}}$ as 
\ba P_{1}^{\rm B} &=& \frac{1}{ (1 + \mu_{\rm E})^{2}}, \non\\
P_{2}^{\rm B} &=& P_{3}^{\rm B}= \frac{\mu_{\rm E}}{(1 + \mu_{\rm E})^{2}},\non\\
P_{4}^{\rm B}&=& \frac{\mu_{\rm E}^2}{(1 + \mu_{\rm E})^{2}} .
\label{PBs}
\ea
Using these expressions, we can write the steady state density matrix of the system at $\bf B$
in terms of $\mu_{\rm E}$.  
It is to be noted that the density matrix is independent of 
$k$ in these limits.
As mentioned before, 
we choose an \emph{energizing} bath which results in a high-entropy state at {\bf B}, or small $\tau_1$,  
or both so that $\rho_{\rm C}=\rho_{\rm B}$.
As shown in Appendix~\ref{appA}
the energy at $\bf B$ ($\mathcal{E}_{\rm B}$), 
and $\bf C$ ($\mathcal{E}_{\rm C}$) can now be written as
\begin{eqnarray}
\mathcal{E}_{\rm B}= L h_1 \frac{\mu_{\rm E}-1}{\mu_{\rm E} + 1}, \nonumber\\
\mathcal{E}_{\rm C}= L h_2 \frac{\mu_{\rm E}-1}{\mu_{\rm E} + 1}.
\end{eqnarray}
Since the decay bath takes the system very close to the ground state
$\mathcal{E}_{\rm D}=-L|h_2|$ for $h_2\ll-1$. 
We write $\mathcal{E}_{\rm A}$ as $\mathcal{E}_{\rm A}^{\rm G}  + \mathcal{E}_{\rm ex, A}$ where
$\mathcal{E}_{\rm A}^{\rm G} $ is the ground state energy corresponding to the Hamiltonian at $\bf A$ and is 
equal to $-Lh_1$. $\mathcal{E}_{\rm ex, A}$ is the excess energy, which will show the Kibble-Zurek scaling.
The work done $\mathcal{W}$ by the system is 
$-(\mathcal{Q}_{\rm in}+\mathcal{Q}_{\rm out})$, which can be simplified 
using the above discussion, and can be written as
\begin{eqnarray}
\mathcal{E}_{\rm ex, A} &=& \mathcal{W}+\frac{2L}{1+\mu_{\rm E}}(\mu_{\rm E}h_1-|h_2|)\nonumber\\
&\propto & \tau_2^{-\frac{\nu d} {\nu z +1}}=\tau_2^{-\frac{1}{2}}
\end{eqnarray}
since $n_{\rm ex} \propto \mathcal{E}_{\rm ex, A}$ 
in the paramagnetic phase, and $\nu=z=1$
for transverse Ising model.
We verify this scaling in Fig. \ref{fig_kz}, which establishes the novel relation \eqref{workkzm} between the work done in a QE and the universal critical exponents of the quantum critical point
crossed.
For numerical
calculations, the initial density matrix is evolved
as per the Eq. (\ref{eq_timstroke1}) when connected to bath,
whereas in the unitary stroke it is simply given by 
\eqref{eq_unitary}
The energies at {\bf A}, {\bf B}, {\bf C}, {\bf D} are calculated
to obtain $\mathcal{Q}_{\rm in}$, $\mathcal{Q}_{\rm out}$, and the work done.
This work done, upto an additive constant,
shows the universal
scaling as shown in Eq. \eqref{workkzm2}, 
and is plotted in Fig. \ref{fig_kz}.
We also show the efficiency of the engine as a 
function of $\tau_2$ in the inset of Fig \ref{fig_kz} 
which approaches a constant value for large $\tau_2$.

To further characterize the performance, we consider the
power as a function of $\tau_2$
in the inset (b) of Fig. \ref{fig_kz}.
As discussed in section \ref{seckzm},
both analytical as well as numerical curves show a peak at
$\tau_2=\tau_{\rm opt} $.
The difference between the 
numerical data and the analytical result is mainly
because the Kibble Zurek scaling, which also appears in the
expression for power, is valid only for large
$\tau_2$ whereas the peak occurs at smaller values.
The analytical and numerical values of efficiency
at maximum power, $\hat \eta$, are respectively given by 0.81 and 0.83, are thus in good agreement.
The figure being a in log-log plot  captures the $1/\tau_2$ behavior of power for large $\tau_2$
which can be explained using Eq. (\ref{eq_powerscaling}).

\subsection{Para-Ferro QE}
\label{tim_num}

\begin{figure}
\includegraphics[height=2.5in]{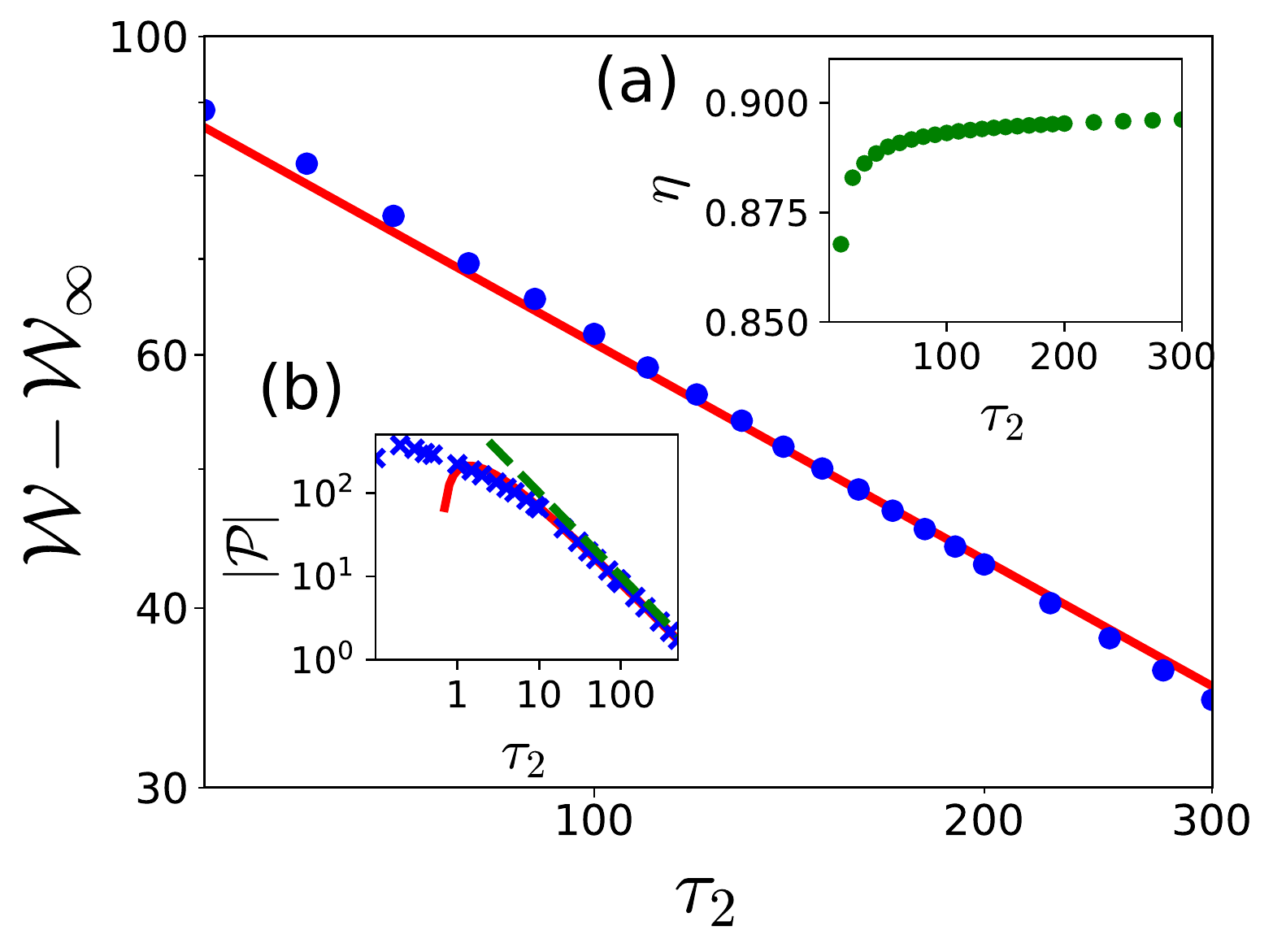}
\caption{{\bf Work output showing universal Kibble-Zurek scaling in Para-Ferro QE.}
The points are the numerical values and red solid line corresponds to 
$\tau_2^{-1/2}$.
Inset (a): Variation of $\eta$ with $\tau_2$.
(b) Variation of $\mathcal{P}$ with
$\tau_2$. The green dashed line corresponds to $1/\tau_2$
scaling, points represent
numerical data and solid line is the analytical expression. The parameters used are: 
$L=100, h_1=10, h_2 = 0, \tau_1=0.01, \mu_{\rm E}'=1, 
\mu_{\rm E}=0.995$ 
with $\mathcal{W}_{\infty}=-899.995$.}
\label{fig_work_para_ferro}
\end{figure}

We realize Para-Ferro QE by considering $h_1 \gg 1$ 
and $0 < h_2 < 1$, such that only the paramagnetic - ferromagnetic critical point is crossed during the unitary strokes. We consider an \emph{energizing} bath of the form shown in Eq. \eqref{eq_timstroke1}, and a \emph{relaxing} bath $\mathcal{B}_{\rm R}$ which
takes the system close to its ground state. Similar to the previous case of Para-Para engine, the work done $\mathcal {W}$,
up to some constant additive factor, will show Kibble-Zurek scaling, as long as the conditions given in Sec. \ref{seckzm} are satisfied. This is presented in Fig. \ref{fig_work_para_ferro}.

\begin{figure}[h]
\centering
\includegraphics[height=2.5in]{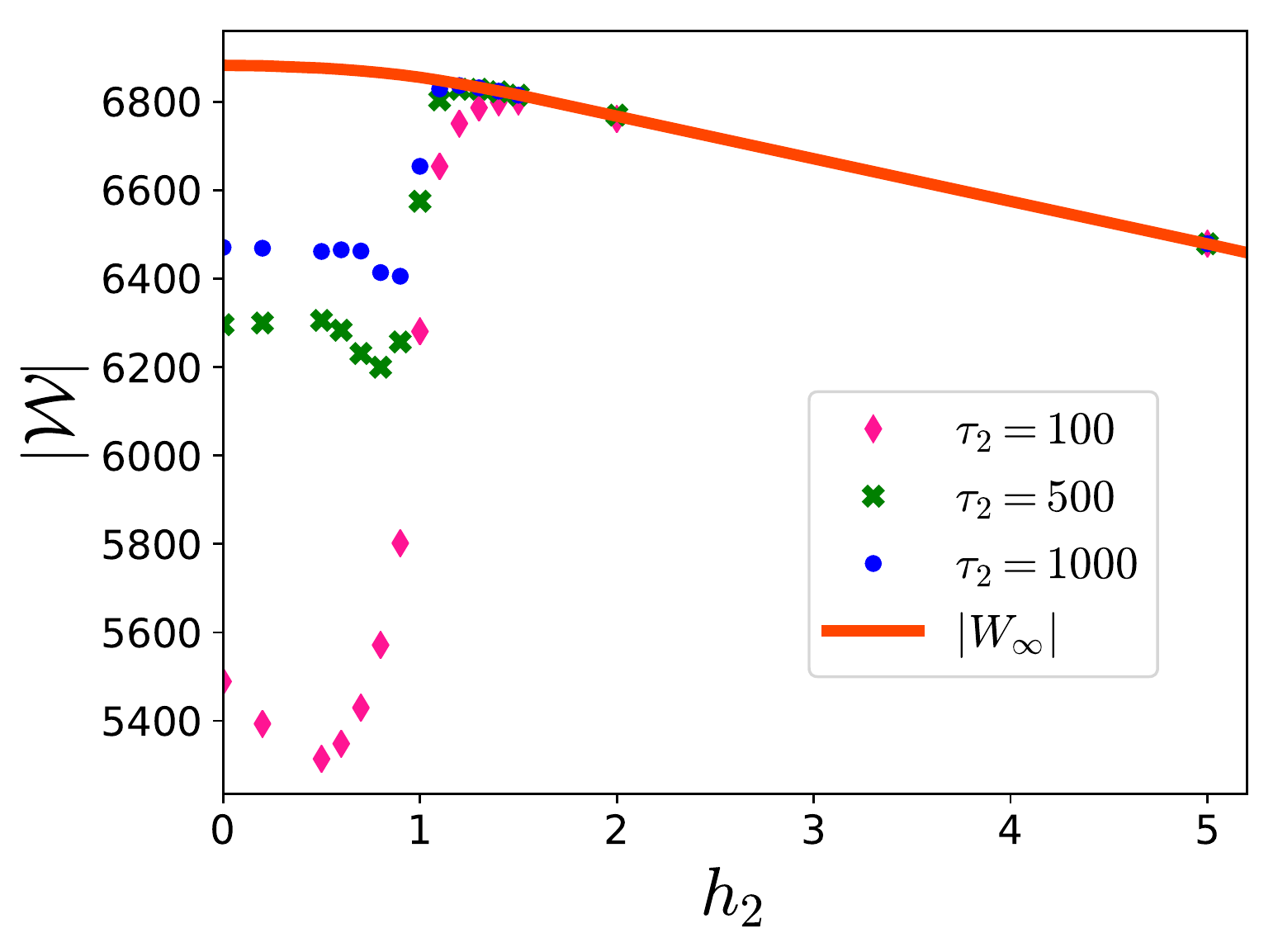}
\caption{{\bf{Work done as a function of $h_2$ for different 
$\tau$ values
along with $W_{\infty}$, as defined in the main text}}. $\mathcal{W}$
improves as $\tau$ increases, or, when adiabaticity is increased.
The parameters used are:$L=100, h_1=70, \tau_1=0.01, \mu_{\rm E}'=1,
\mu_{\rm E}=0.995$.
}
\label{fig_excitation}
\end{figure}

In order to understand the connection between excitations
and engine parameters, we plot below $|W|$ as a function of $h_2$
in Fig. \ref{fig_excitation}
We observe a decrease in power
and work done as the critical point is approached,
which can be attributed
to the excitations produced near the critical point, 
tantamount to quantum friction.
On the other hand, in absence of non-adiabatic excitations 
expected for slow quenches  $\tau_2 \to \infty$ and shortcuts 
to adiabaticity \cite{hartmann19}, driving the quantum
engine across a quantum critical point can boost the
total work output.

\subsection{Critical-Ferro QE}

\begin{figure}
\includegraphics[height=2.5in]{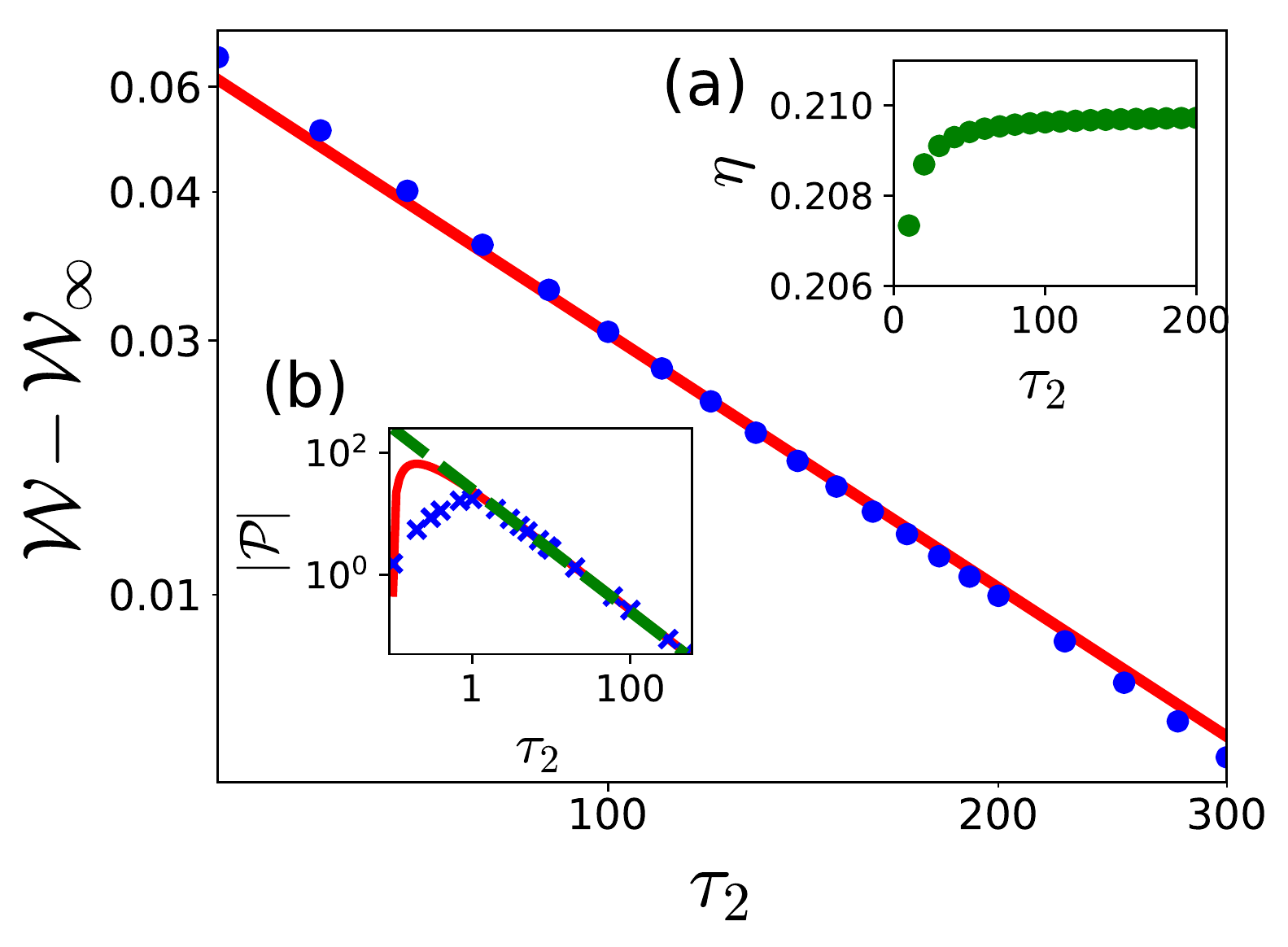}
\caption{{\bf {Work output showing modified Kibble-Zurek
scaling when $h_1$ is close to the critical point.}}
The points are the numerical values and red solid line
corresponds to $\tau_2^{-1}$.
Inset:(a) Variation of $\eta$ with $\tau_2$.
(b) Variation of $\mathcal{P}$ with $\tau_2$
where points represent
numerical data and solid line is the analytical expression.
Also drawn is the green dashed line showing $1/\tau_2$ scaling.
The parameters used are:
$L=100, h_1=0.99, h_2 = 0.0, \tau_1=0.01, \mu_{\rm E}'=1, 
\mu_{\rm E}=0.995$
with $\mathcal{W}_{\infty}=-26.532$.
}
\label{fig_work_crit}
\end{figure}

As described in section \ref{seckzm}, 
we now present the numerical results when $h_1$ is
set close to its critical value of unity. 
During the stroke from {\bf D} to {\bf A} 
the transverse field is linearly varied from $h_2 = 0$ to 
a value close to its critical value of unity. 
As discussed in Ref. \cite{grandi10quench,fei2020work},
the scaling of excess energy gets modified.
Putting $\nu=1$, $z=1$ and $d=1$ in Eq. \eqref{workkzm3}
we get $\tau_2^{-1}$ as shown in Fig. \ref{fig_work_crit},
provided all
the conditions of Sec. \ref{seckzm}  are satisfied.

\subsection{Generalized QE}

Here we present the most general QE without
any restrictions on the relaxing bath, i.e., 
without necessarily  taking the WM to its ground state.
In principle, Generalized QE can take
any bath parameters and $h_1$, $h_2$, provided it works
as an engine, but as we explain below, the analytical
expressions are evaluated under certain conditions.
We focus on
elucidating how the engine parameters change as $h_2$
is varied across the critical point for fixed $h_1$ and other
parameter values.
With $h_1 \gg 1$ and $h_2 > 0$, we choose the \emph{energizing}
bath $\mathcal{B}_{\rm E}$
parameters to be $\mu'_{\rm E} = 1$ and $\mu_{\rm E} < 1$ and
 the \emph{relaxing} bath $\mathcal{B}_{\rm R}$ parameters to 
be $\mu'_{\rm R} = 1$
and $\mu_{\rm R} < \mu_{\rm E}$. This set of parameters related 
to the \emph{relaxing} bath will take the system
to some steady state which is not the ground state at {\bf{D}}.
One can obtain analytical expressions
along the same lines as in the para-para section, 
also presented in Appendix 
but with the condition that $\rho_B=\rho_C$
and $\rho_D=\rho_A$. This is true as long as the critical point
is not crossed or $h_2>1$. The deviation between numerics and analytics
start appearing when $h_2$ approaches the critical point.

\begin{figure}
\includegraphics[height=2.5in]{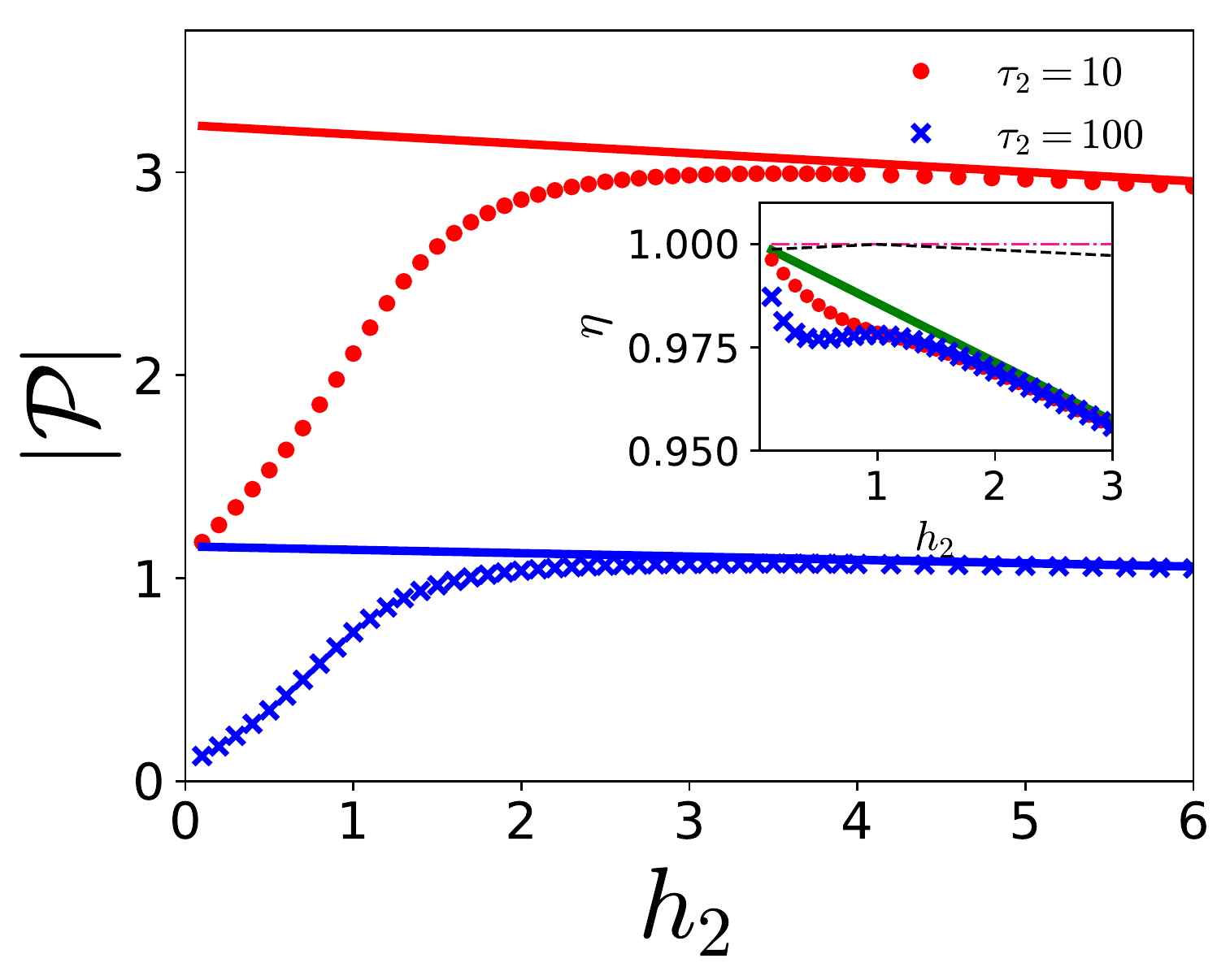}
\caption{{\bf Variation of $|\mathcal{P}|$ as a function of $h_2$ for fixed $h_1$, $\mu_{\rm E}$, $\mu_{\rm R}$,
but for different $\tau_2$ values.} The points correspond to 
numerical values and solid lines to analytical. 
Inset: Variation of $\eta$ and $\eta_{\rm max}$ (black dotted line) 
as a function of $h_2$. $\eta_{\rm max}$ is calculated using 
Eq. (\ref{etamax}) with $\nu=z=1$
for transverse Ising model. 
The maximum value of $\eta_{\rm max}$ is always
less than unity as also confirmed by dashed-dotted line 
corresponding to $\eta=1$.
The other parameters are:
$L=100, h_1=70, \tau_1=0.1, \mu'_{\rm E}=1, \mu_{\rm E}=0.995, \mu'_{\rm R}=1, \mu_{\rm R}=0.95$.}
\label{fig_ph2}
\end{figure}

We shall focus on the behavior of the
QE power output $\mathcal{P}$ for different values of $h_2$.
The final expressions in the limit of large $h_1$ and $h_2$ are:

\ba
\label{EqetaP}
\eta &=& 1 - \frac{h_{2}}{h_{1}} \label{tim_eff},\\
\mathcal{P} &=& -\frac{L (h_{1} - h_{2})}{\tau_{\rm total}} \left(
\frac{\mu_{\rm E}-1}{ \mu_{\rm E} + 1} - \frac{\mu_{\rm R}-1}{\mu_{\rm R} +1}
 \right).
\ea

In Fig. \ref{fig_ph2}, we present the behavior of power $\mathcal{P}$ as a function
of $h_2$ for fixed $h_1$, $\mu_{\rm E}$, $\mu_{\rm R}$ $\tau_1$, and different $\tau_2$
values. Clearly, there is a better
agreement between numerical and analytical values of $\mathcal{P}$ for larger $h_2$.
 Deviations between the two are more pronounced as  the critical point
is approached,  as excitations generated with  the crossing of
the critical point are not included in the analytical calculations (\ref{EqetaP}).
The power $\mathcal{P}$ shows a sharp fall for a QE driven across the phase transition.
This behavior can also be attributed to the excitations produced in the WM close to criticality,
which in turn results in diminishing  $\mathcal{Q}_{\rm in}$, and thus reduce the output power.
We note that  $|0,0\rangle, |1_k,0\rangle, |0,1_{-k}\rangle, |1_k,1_{-k}\rangle$ stop
being  eigenbasis of the WM for small $h_2$.

In the inset of Fig. \ref{fig_ph2}, we present the behavior of efficiency
as a function of $h_2$ for different $\tau_2$ values.
As in the previous case, there is a good match between the
analytical and numerical calculations when away from the QCP, in the paramagnetic phase.
On the other hand, analytical calculations of Sec. \ref{tim_an} fail to explain
the numerical results obtained close to the QCP and in the ordered ferromagnetic phase.
As is expected from Eq. (\ref{tim_eff}), the efficiency is independent of $\tau_2$ when
the operation is confined inside the paramagnetic phase, for large $h_1, h_2$.
However, for a QE driven across a phase transition ($h_1 > 1, ~h_2 < 1$),
as shown in Fig. \ref{fig_ph2} (inset), the results can be expected to
depend non-trivially on $\tau_2$, owing to the dependence of the non-adiabatic excitations
on the rate of driving $\tau_2^{-1}$ across the
QCP \cite{zurek05dynamics, polkovnikov05universal, dutta15quantum}.

\section{Conclusion}
\label{secIV}

We have studied the effect of quantum criticality in quantum thermodynamics, by considering a many-body quantum machine operating close to a phase transition. 
As a WM for the Otto cycle studied here, we have considered interacting Fermions coupled to local dissipative baths, which in the Fourier-transformed space, 
can be treated as non-interacting Fermions coupled to local non-interacting Fermionic dissipative 
baths. This property makes the setup analytically solvable in many regimes. Earlier studies on dynamics of closed many-body systems driven across a quantum critical point have shown the
existence of universal finite-time scaling with the driving speed of different observables, including
 defect density \cite{zurek05dynamics, polkovnikov05universal, mukherjee07quenching} and its fluctuations \cite{delcampo18,Cui19,Bando20}, and
 fidelity susceptibility \cite{grandi10quench, mukherjee11oscillating}, among other examples.  Such finite time scaling can be justified from the diverging length and time scales close to a quantum critical point. 
 In this work,  we have shown for the first time the existence of such universality in quantum thermodynamics close to phase transitions, 
 in the form of Kibble-Zurek scaling \cite{zurek05dynamics, polkovnikov05universal} in the work output, and the 
operation of quantum engines close to criticality.
 Furthermore, we have derived a maximum efficiency bound $\eta_{\rm max}$, which scales with the 
dynamical critical exponent close to quantum criticality, and increases with increasing system size, thus showing the advantage of developing many-body quantum engines.

We have demonstrated these generic results using the model of Ising spin chain in presence of a transverse field. Our analytical and numerical results show that 
the work output inherits a Kibble-Zurek scaling form, up to an additive constant, for a quantum engine driven across quantum critical points ($h_1 \gg 1$, $h_2 \ll -1$ or $h_1 \gg 1$, $-1 < h_2 < 1$).  
By contrast, for a quantum engine confined to the paramagnetic phase,
the power attains a maxima close to the QCP ($h_1 \gg 1$, $h_2 > 1$), rapidly decreasing once the WM
approaches the QCP ($h_1 \gg 1$, $h_2 \to 1^{+}$), diminishing close to zero when the 
efficiency is maximum. The loss of power in this case can be attributed to the generation of 
excitations close to quantum criticality.

While we have mainly focussed on Fermionic baths, our results  can be expected to be valid for other kind of baths, as long as the conditions stated in Sec. \ref{seckzm} are satisfied.
We note that in this case the relaxing bath would 
be a thermal bath at absolute zero temperature, such that the 
WM reaches close to its ground state at the end of the non-unitary 
stroke {\bf C} to {\bf D}. Consequently, the efficiency of the 
quantum heat engine would be bounded by the Carnot limit of 
maximum efficiency, which in this case reduces to the trivial 
result $\eta \leq 1$.
In addition, by considering thermal instead of Fermionic bath, our setting can be readily adapted to the characterization of quantum refrigerators. 

The class of  quantum  machines studied here 
provides an opportunity to scale up  quantum devices to the macroscopic regime, with a complete understanding of their performance. 
Experimental implementations can be envisioned  in an optical lattice setup \cite{schreiber15observation}.
Our results should also be of relevance to
the scaling of quantum machines using trapped ion chains as a working medium \cite{Rossnagel16,Maslennikov17,vonLindenfels19} in which  a quantum Ising chain 
can be emulated \cite{Friedenauer08,Zhang17,Bernien17} and in which universal critical dynamics has been studied \cite{delcampo10,Silvi13,Silvi16},  with experiments reported to date probing it in the classical regime \cite{EH13,Ulm13,Pyka13}.   
Nuclear magnetic resonance experiments  and 
nitrogen vacancy centers offer alternative platforms in which 
the  quantum engines reported to date \cite{Peterson18,Klatzow19}  
can be scaled up  considering quantum  critical spin systems as 
working substance. 
Beyond specific implementations, our results advance  the study of universal critical phenomena in quantum thermodynamics.

\begin{acknowledgments}
It is a pleasure to thank Fernando J. G\'omez-Ruiz for useful discussions and comments on the manuscript.
UD acknowledges DST, India for INSPIRE Research grant. UD also acknowledge the hospitality of the
Donostia International Physics Center, Spain, and IISER Berhampur, India, during her visits.
VM acknowledges Amit Dutta for fruitful discussions, SERB, India for Start-up Research Grant SRG/2019/000411 and IISER Berhampur for Seed grant.
This work is further  supported by ID2019-109007GA-I00.

\end{acknowledgments}

\appendix

\section{Dynamics of the Working Medium}
\label{appA}

\textbf{Para - Para engine}\\

The density matrix at $\bf B$ for each $k$ mode takes the form given in Eq. (\ref{rho_B}) so that the energy is calculated as
$\mathcal{E}_{\rm B} = \sum_k \text{Tr}[H(h_1,k)\rho_{\rm B}]$.
Therefore for each $k$ mode,

$H \rho_{\rm B} =$
\begin{equation} 
\begin{split}
 \footnotesize{
\begin{bmatrix} 2(h_{1} + \cos k) & 0 & 0 & 2
\sin k \\ 0 & 0 & 0 & 0\\ 0 & 0 & 0 & 0\\ 2 \sin k & 0 & 0 & -2(h_{1} + \cos k)
\end{bmatrix} 
\begin{bmatrix} P_{4}^{\rm B} & 0 & 0 & 0 \\ 0 & P_{3}^{\rm B} & 0 & 0\\
0 & 0 & P_{2}^{\rm B} & 0\\ 0 & 0 & 0 & P_{1}^{\rm B} 
\end{bmatrix}}
\end{split} 
\end{equation}
which gives \begin{align} \text{Tr}[H\rho_{\rm B}] &= 2(h_{1} + \cos k) (P_{4}^{\rm B}
- P_{1}^{\rm B})\\ &= 2(h_{1} + \cos k) \left( \frac{\mu_{\rm E} - 1}{ \mu_{\rm E} + 1} \right).\end{align} 
For a system of size $L$, there are $L/2$ positive $k$ modes so that
\ba
\mathcal{E}_{\rm B} &=& \sum_k 2(h_{1} + \cos k) \left( \frac{\mu_{\rm E} - 1}{ \mu_{\rm E} + 1} \right)  \nonumber\\
&=& L h_{1} \left(\frac{\mu_{\rm E} - 1}{ \mu_{\rm E} + 1} \right).\label{E_B} 
\ea
Since the Hamiltonian is changed suddenly
(small $\tau_1$) from $h_1$ to $h_2$, the density matrix 
is not able to evolve resulting to $\rho_C = \rho_B$ and thus 
  energy at $\bf C$ $(\mathcal{E}_{\rm C})$ is 
\ba 
\mathcal{E}_{\rm C} &=& \sum_k 2(h_{2} + \cos k) \left( \frac{\mu_{\rm E} - 1}{ \mu_{\rm E} + 1} \right) \nonumber\\
&=& L h_{2} \left(\frac{\mu_{\rm E} - 1}{ \mu_{\rm E} + 1} \right). 
\label{E_C}
\ea
Since $\bf D$ is in the ground state, we have 
$\mathcal{E}_{\rm D} = -L|h_{2}|$.
We write energy at $\bf A$ to be $\mathcal{E}_{\rm A} = \mathcal{E}_{\rm A}^{\rm G} + \mathcal{E}_{\rm ex, A}$ with $\mathcal{E}_{\rm A}^{\rm G}  = -L h_1$. This gives
\begin{align}
\mathcal{Q}_{\rm in} &= \mathcal{E}_{\rm B} - \mathcal{E}_{\rm A}\\
&= \frac{2L \mu_{\rm E}h_{1}}{\mu_{\rm E} + 1} - \mathcal{E}_{\rm ex, A}
\end{align}

and
\begin{eqnarray}
\mathcal{Q}_{\rm out} &=& \mathcal{E}_{\rm D} - \mathcal{E}_{\rm C} \nonumber\\
&=& \frac{-2L |h_{2}|}{\mu_{\rm E} + 1}.
\end{eqnarray}
The work done, $\mathcal{W} = -(\mathcal{Q}_{\rm in} + \mathcal{Q}_{\rm out})$, is thus
\begin{equation}
\mathcal{W} = - \left( \frac{2L}{\mu_{\rm E} + 1} (\mu_{\rm E}h_{1} 
- |h_{2}|) - \mathcal{E}_{\rm ex, A} \right)
\end{equation}
or equivalently
\begin{equation}
\mathcal{W} + \frac{2L}{\mu_{\rm E} + 1} (\mu_{\rm E}h_{1} 
- |h_{2}|) \propto \tau_2^{\frac{-\nu d}{1 + \nu z}}.
\end{equation}

\textbf{Generalized QE}\\

Clearly, there is no change in  $\mathcal{E}_{\rm B}$ 
so that it is given by Eq. (\ref{E_B}). For large
$h_2$ with $h_2<h_1$, there will not be any population
change in B to C, and hence the density matrix $\rho_{\rm B}$ will be same
as $\rho_{\rm C}$ so that $\mathcal{E}_{\rm C}$ is also given by 
Eq. (\ref{E_C}).

The energy at $\bf D$ would be different since the \emph{relaxing}
bath parameters are so chosen that it need not take the system
to the ground state. It can be calculated as follows: 
\begin{align}
\mathcal{E}_{\rm D} &= \sum_k \text{Tr}[H(h_2,k)\rho_{\rm D}]\\
&= \sum_k 2(h_{2} + \cos k) (P_4^{\rm D} - P_1^{\rm D})\\
&= \sum_k 2(h_{2} + \cos k) \left( \frac{\mu_{\rm R} - 1}{ \mu_{\rm R} + 1} \right)\\
&= Lh_2 \left( \frac{\mu_{\rm R} - 1}{ \mu_{\rm R} + 1} \right).
\end{align}
Here, $\rho_k^{\rm D}$ would be similar to $\rho_k^{\rm B}$ as given in 
Eq. (\ref{rho_B})
with $\mu_{\rm E}$ replaced by $\mu_{\rm R}$.
Similar calculations give 
\begin{align}
\mathcal{E}_{\rm A} &= \sum_k \text{Tr}[H(h_1,k)\rho_{\rm A}]\\
&= Lh_1 \left( \frac{\mu_{\rm R} - 1}{ \mu_{\rm R} + 1} \right).
\end{align}
Now, the  $\mathcal{Q}_{\rm in}$ and  $\mathcal{Q}_{\rm out}$ for each $k$ mode is 
\begin{eqnarray}
\mathcal{Q}_{\rm in}(k) &= 2(h_{1} + \cos k)
\left(\frac{\mu_{\rm E}-1}{ \mu_{\rm E} + 1} - \frac{\mu_{\rm R}-1}{\mu_{\rm R}+1} \right),\\ 
\mathcal{Q}_{\rm out}(k) &= - 2(h_{2} + \cos k) \left(\frac{\mu_{\rm E}-1}{ \mu_{\rm E} + 1} - \frac{\mu_{\rm R}-1}{\mu_{\rm R} +1} \right).
\end{eqnarray}

Let 
\begin{equation} 
\frac{\mu_{\rm E}-1}{ \mu_{\rm E} + 1} - \frac{\mu_{\rm R}-1}{\mu_{\rm R} +1} = \alpha  .
\end{equation} 
Efficiency of the total system can be calculated using

\begin{align} \eta &=
\frac{\sum_{k} \mathcal{Q}_{\rm in}(k) + \sum_{k} \mathcal{Q}_{\rm out}(k)}{\sum_{k} \mathcal{Q}_{\rm in}(k)}\\
&=
\frac{2 \alpha  \sum_{k}[(h_{1} + \cos k) - (h_{2} + \cos k)]} {2 \alpha \sum_{k}(h_{1} + \cos k)}\\ 
&=
\frac{\frac{L}{2}(h_{1} - h_{2})}{\frac{L}{2} h_{1} + \underbrace{\Sigma_{k} \cos k}}_{=0} \\ &= 1 -
\frac{h_{2}}{h_{1}}.
\end{align} 
Power for the total system is defined as
\begin{align} 
\mathcal{P} &= \frac{\text{net work done by the
system}}{\text{total cycle time}}\\ 
&= -\frac{2 \alpha \sum_{k}[(h_{1} + \cos k)  -
(h_{2} + \cos k)]}{\tau_{\rm total}}\\
 &= -\frac{2 \alpha (\frac{L}{2}) (h_{1} -
h_{2})}{\tau_{\rm total}}\\ 
&= -\frac{L (h_{1} - h_{2})}{\tau_{\rm total}} \left(
\frac{\mu_{\rm E}-1}{ \mu_{\rm E} + 1} - \frac{\mu_{\rm R}-1}{\mu_{\rm R} +1}
 \right) .
\end{align}

\end{document}